\documentclass[a4paper,11pt]{article}
\pdfoutput=1
\usepackage{jheppub}
\hypersetup{
  colorlinks=false,
  pdfborder={0 0 1},
  linkbordercolor={0 0 1},
  citebordercolor={0 0 1},
  urlbordercolor={0 0 1}
}
\usepackage{cancel}
\usepackage{subcaption}
\usepackage{mathrsfs}
\usepackage{amsmath}   
\usepackage{feynmp-auto,expdlist}
\usepackage{float}

\newcommand{\beq}[1]{\begin{equation}\label{#1}}
\newcommand{\eeq}{\end{equation}}
\newcommand{\bea}[1]{\begin{eqnarray}\label{#1}}
\newcommand{\eea}{\end{eqnarray}}

\title{Probing quantum chaos near a wormhole throat with a circular string}
 \author{Ai-chen Li $^{a,b}$, Xin-Fei Li $^{b}$, Xuanting Ji $^{c}$}
 \affiliation[a]{Department of Physics, Institute of Fundamental Physics and Quantum Technology, Ningbo University, 818 Fenghua Road, Ningbo, 315211, Zhejiang, China}
 \affiliation[b]{School of Science, Guangxi University of Science and Technology, 545026 Liuzhou, China} 
\affiliation[c]{Department of Applied Physics, College of Science, China Agricultural University, Beijing 100083, China}

\emailAdd{alexkenlee@163.com}
\emailAdd{xfli@gxust.edu.cn}
\emailAdd{jixuanting@cau.edu.cn}

\abstract{We investigate whether quantum fluctuations of a circular probe string develop a quantum-chaotic response while traversing a wormhole throat. The classical circular-string embedding is periodic and radially stable, but its two physical transverse polarizations experience time-dependent tidal potentials. Expanding the world-sheet action to quadratic order, we canonically quantize these modes and construct out-of-time-ordered correlator(OTOC) amplitudes from their unequal-time commutators. For the Ellis--Bronnikov wormhole, both polarizations exhibit finite intervals of approximately exponential OTOC growth associated with the first throat passage. The corresponding dimensionless rate measured with respect to physical time is positive over the parameter range studied and generally decreases as the probe energy is increased relative to the throat scale. In the global-monopole extension, increasing the solid-angle deficit narrows the band of locally amplifiable modes and suppresses the extracted rates; a sufficiently strong defect can nearly quench the radial signal, while the angular channel retains a polarization-dependent non-monotonic structure when the energy-to-throat-scale ratio is small. These quantities characterize finite-time dynamical sensitivity in the Gaussian fluctuation sector and should not be identified with asymptotic many-body chaos or a thermodynamic phase transition. Although the numerical analysis uses two representative wormhole geometries, the construction depends only on covariant world-sheet fluctuations and real-time commutators. It therefore provides a transferable, non-holographic framework for applying quantum-chaos diagnostics directly to quantum probes in curved spacetimes.}

\begin{document} 
\maketitle
\flushbottom

\section{Introduction}

The study of chaos in black-hole spacetimes first developed through relativistic trajectory dynamics. Although geodesic motion in an isolated Schwarzschild or Kerr geometry is integrable, spin--curvature coupling and multi-centre gravitational fields were shown to generate non-integrable motion, chaotic scattering and fractal basin structures \cite{Suzuki:1996SpinningChaos,Aguirregabiria:1996ChaoticScattering}. Near-horizon shock-wave calculations subsequently shifted the emphasis towards universal sensitivity to initial perturbations and rapid information scrambling \cite{Shenker:2013pqa,Maldacena:2015waa}. A complementary local mechanism was identified in Ref.~\cite{Hashimoto:2016dfz}: a particle supported by an external force sufficiently close to a non-extremal, spherically symmetric horizon sits near an unstable potential maximum, with a Lyapunov exponent fixed by the surface gravity independently of the detailed force and probe mass. Representative developments have since examined this mechanism in generic static and rotating horizons \cite{Dalui:2018HorizonChaos}, tested the associated bound in charged and nonlinearly coupled black-hole backgrounds \cite{Zhao:2018ChaosBound,Lei:2020QuasitopologicalChaos}, and related horizon-induced instability to inaffinity for both black-hole and cosmological horizons \cite{Giataganas:2021HorizonsChaos}. More recently, near-ring OTOCs have reproduced the instability exponent of equatorial circular null orbits in generalized Kerr geometries and, together with the temperature measured by a string probe, were found to saturate the quantum chaos bound at the photon ring \cite{Giataganas:2026PhotonRings}. The corresponding question for a regular traversable wormhole throat remains much less developed. Existing connections between wormholes and chaos have largely been formulated through coupled Sachdev--Ye--Kitaev (SYK) systems, holographic gauge/gravity models or Euclidean spectral correlations \cite{Nosaka:2020nuk,Cubrovic:2021puw,Haehl:2023EuclideanWormholes}, rather than through the local dynamics of a probe traversing a Lorentzian throat.

The recent analysis of Ref.~\cite{Zhao:2026ghk} provides a notable exception: for an externally confined particle, the one-dimensional radial sector is stable near the throat, whereas sufficiently energetic two-dimensional motion develops chaotic regions while retaining surviving KAM islands. The absence of chaos in the linear radial problem has a simple origin. At a regular static throat the lapse remains finite and the effective potential is smooth; expanding about a stable trajectory to quadratic order produces linear oscillator-type equations, without the nonlinear mode coupling and phase-space folding required for sustained chaotic dynamics. Time-dependent coefficients may nevertheless generate transient tachyonic or parametric amplification, so this observation does not exclude finite-time sensitivity in the fluctuation sector. Establishing whether such sensitivity can be generated by the throat is therefore physically significant: it isolates the roles of tidal geometry and nontrivial topology without a horizon, sharpens dynamical distinctions between wormholes and black-hole mimickers, and provides a controlled arena in which traversability, information propagation and strong-field instability can be studied together.

In parallel, out-of-time-ordered correlators (OTOCs) have become one of the principal diagnostics of quantum chaos and information scrambling across quantum physics \cite{GarciaMata:2022OTOCReview}. Originating in the analysis of disordered superconductors \cite{Larkin:1969Quasiclassical}, an OTOC or its squared-commutator form measures the failure of initially commuting operators to remain compatible under Heisenberg evolution and thereby resolves the growth and spatial propagation of quantum perturbations. In non-gravitational systems, OTOCs have quantified early-time sensitivity and Ehrenfest-time effects in few-body quantum maps and the kicked rotor \cite{Rozenbaum:2016OTOC,Hashimoto:2017oit}; distinguished ballistic, diffusive and logarithmic information fronts in interacting spin chains and many-body-localized phases \cite{Luitz:2017jvm,Huang:2016MBLOTOC}; characterized operator fronts and butterfly velocities in random unitary circuits \cite{Nahum:2017OperatorSpreading}; and diagnosed scrambling in periodically driven systems \cite{Nizami:2020agu}. Recent developments have further related OTOC propagation to the compressibility of disordered Floquet random circuits and introduced operator-shadow and Bell-sampling protocols, together with query-complexity bounds, for learning large families of dynamical correlators \cite{DeFranco:2026FloquetCompressibility,Chiew:2026OperatorShadows}. They also admit a thermodynamic interpretation in terms of work statistics and nonlinear response \cite{Campisi:2016Thermodynamics,Tsuji:2016OTOFDT}, and have been measured through time-reversal protocols in trapped-ion magnets and nuclear-magnetic-resonance quantum simulators \cite{Gaerttner:2016IonOTOC,Li:2016NMROTOC}. In gravitational settings, OTOCs organize the shock-wave description of black-hole scrambling and the chaos bound \cite{Shenker:2013pqa,Roberts:2014Localized,Maldacena:2015waa}, extend to far-from-equilibrium states dual to collapsing BTZ--Vaidya geometries \cite{Balasubramanian:2019QuenchChaos}, and probe de Sitter expansion, stochastic inflation, reheating and squeezed cosmological perturbations \cite{Choudhury:2018Cosmology,Aalsma:2020DeSitter,Haque:2020SqueezedOTOC}. Recent de Sitter analyses have sharpened this connection through observer-dressed eikonal OTOCs, shockwave-induced anti-scrambling algebras and static-patch trace obstructions associated with observer recoil and gravitational backreaction \cite{Milekhin:2026dSOTOC,Cui:2026AntiScrambling,Chen:2026NegativeShocks}. Stringy corrections to holographic scattering and open-string fluctuations on probe branes furnish further gravitational applications \cite{Shenker:2014Stringy,Banerjee:2018Strings}. By comparison, direct OTOC studies of material compact stars and other horizonless ultracompact geometries remain comparatively sparse. Across these examples, exponential growth can encode local instability, operator spreading or squeezing and must therefore be interpreted together with the state, time window and degrees of freedom being probed \cite{Xu:2019lhc}; subject to this qualification, OTOCs provide an unusually versatile bridge between dynamical sensitivity, scrambling and quantum many-body evolution.

The purpose of this work is to use OTOCs to test whether quantum fluctuations of a cicular string in a wormhole spacetime, especially during a passage through the throat, can develop a quantum-chaotic response. Our starting point is the framework developed in Refs.~\cite{Li:2026rut,Li:2026oxy}. A circular string is embedded in a spherically symmetric geometry, and the Polyakov action is expanded covariantly to quadratic order in normal deformations of the world sheet, yielding the two physical transverse polarizations. The resulting Gaussian theory admits a mode-by-mode canonical quantization: Wronskian-normalized mode functions preserve the equal-time commutators, while a reference Fock vacuum is specified at a distant point where the string lies outside the strong-curvature region. Here we extend this construction from particle production, entanglement and complexity observables to the unequal-time commutators that determine the OTOC amplitudes. We apply it first to the Ellis--Bronnikov wormhole and then to its global-monopole extension, for which the solid-angle deficit supplies an additional geometric control parameter. This local-probe perspective is complementary to direct non-holographic studies of quantum fields and quantum backreaction in wormhole geometries, which have focused on vacuum polarization, renormalized stress tensors or the quantum support and stability of the throat \cite{Popov:2000WormholeVacuum,Bezerra:2010WormholeVacuum,Maldacena:2018FourDimensions,Mehulic:2026Backreaction}. In contrast, the observable followed here is the real-time operator growth of a quantized probe as it crosses the throat. Although the numerical results are obtained in two specific backgrounds, the method requires neither a holographic dictionary nor an asymptotic AdS boundary: it is formulated in terms of the embedding geometry, the covariant transverse action and the real-time evolution of its mode functions. It can therefore be transferred to other traversable wormholes and, more generally, to horizonless strong-gravity geometries. The present models serve as a minimal proof of principle. We find that the classically stable circular trajectory can coexist with finite intervals of throat-associated OTOC growth in both transverse channels; the accumulated physical-time rate decreases as the probe energy increases relative to the throat scale, while a larger monopole deficit generally suppresses it and can nearly quench the radial signal. These results identify the throat as a localized source of finite-time dynamical sensitivity and the deficit angle as a polarization-dependent control parameter. At the same time, because the calculation remains quadratic, the extracted rates characterize the Gaussian fluctuation sector rather than asymptotic chaos or thermalization in a fully interacting string theory.

The remainder of the paper is organized as follows. In Sec.~\ref{ClaStringSta} we derive the periodic circular-string trajectory in the global-monopole wormhole geometry and establish the radial stability of its classical circular sector. Section~\ref{SemiQuantizeStringQuadra} develops the covariant quadratic action for the two physical transverse fluctuations, performs their canonical quantization and expresses the radial and angular OTOC amplitudes in terms of unequal-time mode-function commutators. Section~\ref{NumAnaOTOC} contains the numerical analysis. Section~\ref{PureEBWHCase} treats the Ellis--Bronnikov limit and extracts the first-passage effective Lyapunov rates, while Sec.~\ref{TopoGMWHCase} determines how the solid-angle deficit modifies the effective frequencies, the amplifiable mode band and the OTOC growth. The independent checks in Sec.~\ref{NumericalValidation} test Wronskian preservation, mode-cutoff convergence, fit-window stability and the numerical uncertainty of the extracted rates. We summarize the physical implications, limitations and possible extensions in Sec.~\ref{ConcluDiscuss}.

\section{Classical circular-string dynamics and radial stability \label{ClaStringSta}}

We first determine the classical circular-string trajectory and its radial stability, providing the background for the fluctuation analysis below. The dynamics follows from the Polyakov action,
\begin{align}
\label{BackStringAction}
&S_{(0)}=\frac{-1}{4\pi\alpha^{\prime}}\int d\tau d\sigma\sqrt{-h}h^{\mathtt{A}\mathtt{B}}G_{\mathtt{A}\mathtt{B}}~,
\end{align}where $h_{\mathtt{A}\mathtt{B}}$ is the intrinsic world-sheet metric and $G_{\mathtt{A}\mathtt{B}}$ is the metric induced by the spacetime embedding,
\begin{align}
\label{InduceMetric}
&G_{\mathtt{A}\mathtt{B}}=g_{\mu\nu}\frac{\partial x^{\mu}}{\partial\xi^{\mathtt{A}}}\frac{\partial x^{\nu}}{\partial\xi^{\mathtt{B}}}\,,\,\xi^{\mathtt{A}}=\{\tau,\sigma\}\,,\,x^{\mu}=\{t,r,\theta,\phi\}~,
\end{align}Variation of \eqref{BackStringAction} with respect to the world-sheet metric $h_{\mathtt{A}\mathtt{B}}$ and the embedding coordinates $x^\mu(\xi)$ yields the embedding equations of motion and the vanishing of the world-sheet energy--momentum tensor,
\begin{align}
\label{EOMsBackEmbedding}
0&=h^{\mathtt{A}\mathtt{B}}(\nabla_{\mathtt{A}}\nabla_{\mathtt{B}}x^{\mu}+\Gamma_{\alpha\beta}^{\mu}\frac{\partial x^{\alpha}}{\partial\xi^{\mathtt{A}}}\frac{\partial x^{\beta}}{\partial\xi^{\mathtt{B}}})~,\\
\label{RestrictBackEmbedding}
0&=\frac{1}{2}h_{\mathtt{A}\mathtt{B}}h^{\mathtt{A}_{1}\mathtt{B}_{1}}G_{\mathtt{A}_{1}\mathtt{B}_{1}}-G_{\mathtt{A}\mathtt{B}}~.
\end{align}Using world-sheet diffeomorphisms and Weyl invariance, we work in conformal gauge, $h_{\mathtt{A}\mathtt{B}}=\eta_{\mathtt{A}\mathtt{B}}$. We consider a circular probe string in the static, spherically symmetric global-monopole wormhole geometry \cite{Li:2026rut},
\begin{align}
\label{NormalSphericalCoordinates}
&ds^{2}=-dt^{2}+\frac{dr^{2}}{\alpha_{0}^{2}(1-\frac{r_{0}^{2}}{r^{2}})}+r^{2}(d\theta^{2}+\sin^{2}\theta d\phi^{2})\,,\,r_{0}=\kappa\eta L_{0}\,,\,\alpha_{0}^{2}\!=\!1\!-\!\kappa^{2}\eta^{2}.
\end{align}Here $\eta$ is the symmetry-breaking scale associated with the topological defect. The geometry is ultrastatic; for a nonzero solid-angle deficit, its asymptotic regions are conical rather than asymptotically Minkowski. To describe a passage through the throat without the coordinate degeneracy of the areal radius, we introduce the proper radial coordinate $\varrho=\sqrt{r^{2}-r_{0}^{2}}/\alpha_{0}$ on one exterior branch and continue it with the opposite sign on the other. The two regions are then covered by the regular metric
\begin{align}
\label{ProCoordiMonoWH}
&ds^{2}=-dt^{2}+d\varrho^{2}+(\alpha_{0}^{2}\varrho^{2}+r_{0}^{2})(d\theta^{2}+\sin^{2}\theta d\phi^{2}).
\end{align}A circular string lying in the equatorial plane, with its radius allowed to vary in time, is described by the embedding
\begin{align}
\label{BackStringEmbed}
&x^{\mu}(\tau,\sigma):t=t(\tau)\,,\,\varrho=\varrho(\tau)\,,\,\theta=\pi/2\,,\,\phi=\sigma~.
\end{align}Substituting the metric \eqref{ProCoordiMonoWH} and the ansatz \eqref{BackStringEmbed} into the embedding equations \eqref{EOMsBackEmbedding} and the Virasoro constraints \eqref{RestrictBackEmbedding} gives
\begin{align}
\label{EOMsComponent}
&0=\ddot{t}(\tau)=\ddot{\varrho}(\tau)+\alpha_{0}^{2}\varrho(\tau),\\
\label{ViraSoroComponent}
&0=\dot{\varrho}(\tau)^{2}-\dot{t}(\tau)^{2}+\alpha_{0}^{2}\varrho(\tau)^{2}+r_{0}^{2},
\end{align}For $\alpha_0>0$ and $\kappa^{2}E>r_0$, the solutions describe nondegenerate throat-crossing trajectories. Direct integration gives
\begin{align}
\label{BackStringTime}
\bar{t}(\tau)&=\kappa^{2}E\tau+t_0,\\
\label{BackStringProperRadial}
\bar{\varrho}(\tau)&=\frac{\sqrt{\kappa^{4}E^{2}-r_{0}^{2}}}{\alpha_{0}}\cos\left(\alpha_{0}(\tau-\tau_{0})\right).
\end{align}Substitution of \eqref{BackStringTime} into the Virasoro constraint \eqref{ViraSoroComponent} expresses the radial dynamics in terms of an effective potential,
\begin{align}
&\dot{\varrho}(\tau)^{2}+V_{\text{eff}}\left(\varrho(\tau)\right)=\kappa^{4}E^{2}\,,\,V_{\text{eff}}(\varrho)=\alpha_{0}^{2}\varrho^{2}+r_{0}^{2}.
\end{align}The effective potential has a regular minimum at the throat, where $\varrho(r_0)=0$. Its local restoring character can be examined through a small radial displacement, $\varrho(\tau)\simeq 0+\delta\varrho(\tau)$, whose linearized equation of motion is
\begin{align}
\ddot{\delta\varrho}+\frac{1}{2}V_{\text{eff}}^{\prime\prime}(0)\delta\varrho=0,\qquad
V_{\text{eff}}^{\prime\prime}(0)=2\alpha_0^2>0.
\end{align}
The solutions are oscillatory. Since the exact radial equation \eqref{EOMsComponent} is linear, the same equation also governs the separation between neighboring circular trajectories, including those crossing the throat. The circular ansatz thus defines an integrable radial sector with no exponentially growing variational mode. This differs from the mechanism identified by Hashimoto and Tanahashi \cite{Hashimoto:2016dfz}: an externally supported particle near a nonextremal black-hole horizon encounters an unstable potential maximum, whose inverted-oscillator dynamics supplies a local instability that can seed chaos when coupled to additional degrees of freedom. Here the lapse is constant, the throat is regular in proper distance, and string tension produces a potential minimum rather than such a barrier. The horizon-based mechanism is therefore absent. Nor does the connection to the thermal chaos bound \cite{Maldacena:2015waa}, discussed in Ref.~\cite{Hashimoto:2016dfz}, provide a universal constraint for this horizonless probe problem without the corresponding thermal assumptions. Importantly, this conclusion concerns radial variations within the circular ansatz, not arbitrary deformations of the string or all particle motion in wormhole backgrounds. Indeed, the externally confined wormhole particle studied in Ref.~\cite{Zhao:2026ghk} is radially stable at linear order but can exhibit nonlinear chaos once a second dynamical coordinate and sufficiently high energies are included.

The trajectory \eqref{BackStringProperRadial} distinguishes two physically different events: a passage through the throat $\bar{\varrho}=0$, connecting the two exterior regions of the same wormhole spacetime, and a turning point at maximal proper distance from the throat, where the radial velocity vanishes. Their world-sheet times are, respectively,
\begin{align}
&\tau_{\text{throat}}=\frac{(1+2n)\pi}{2\alpha_{0}}+\tau_{0},~~\tau_{\text{turning}}=\frac{n\pi}{\alpha_{0}}+\tau_{0},
\end{align}where $n\in\mathbb{Z}$ labels successive events along the orbit. At $\varrho(\tau_{\text{throat}})$, the signed proper radial coordinate changes sign: the string passes smoothly from one exterior region to the other at $\tau=\tau_{\text{throat}}$, with a continuous and nonzero proper radial velocity. By contrast, at each $\varrho(\tau_{\text{turning}})$ the proper radial velocity vanishes and reverses sign, so the string remains in the same exterior region and starts moving back towards the throat. Although the areal radius is stationary at the throat because it reaches its minimum there, this does not represent a reversal of the motion in proper distance. The circular string therefore executes periodic motion between the two exterior turning points, crossing the throat twice per complete orbit. This regular background raises the question that motivates the remainder of the paper: can the quantized transverse fluctuations exhibit the exponential OTOC growth associated with quantum chaos near the throat, even though the circular radial sector has no growing mode? This particular fluctuation problem is not addressed by the classical particle analysis of Ref.~\cite{Hashimoto:2016dfz} or its near-throat counterpart \cite{Zhao:2026ghk}. A circular string is a useful probe precisely because it retains spatially resolved internal modes that respond to tidal curvature and to the extrinsic geometry of the embedding, while its background motion remains analytically tractable. This choice also addresses the finite-size world-sheet effects highlighted in the outlook of Ref.~\cite{Hashimoto:2016dfz}. In the probe and quadratic approximations, the physical transverse sector admits a covariant description in terms of world-sheet fields \cite{Larsen:1993mx,Garriga:1991ts,Guven:1993ex}, allowing a mode-by-mode canonical treatment and a direct construction of unequal-time commutators. The advantage is therefore a controlled description of an extended probe, rather than a general simplification relative to worldline quantization. We now develop this semiclassical fluctuation theory about the circular trajectory, without assuming the full quantum consistency of a fundamental string theory in the given background.

\section{Quantum chaos from quantized string fluctuations \label{SemiQuantizeStringQuadra}}

\subsection{Quadratic action for transverse string fluctuations}

The absence of an exponentially growing classical mode does not preclude quantum scrambling in the fluctuation sector. We therefore expand the circular string around the background solution in a covariant basis adapted to the world sheet. The physical perturbations are transverse to the world sheet, and hence $\delta x^{\mu}$ is projected onto unit normal vectors associated with the embedding $\bar{x}^{\mu}(\tau,\sigma)$. The general construction of the normal frame and the corresponding geometric data may be found in \cite{Larsen:1993mx,Larsen:1998sh,Garriga:1991ts,Guven:1993ex}. For a circular string in the spherically symmetric wormhole geometry \eqref{ProCoordiMonoWH}, the normal bundle is two-dimensional. We choose its two polarizations to be the radial direction $\varrho$ and the polar direction $\theta$. Denoting the polarization index by $\mathtt{i}=\varrho,\theta$, the transverse fluctuation is decomposed as
\begin{align}
&\delta x^{\mu}\left(\tau,\sigma\right)=\bar{n}_{\left(\varrho\right)}^{\mu}\left(\tau\right)\Phi^{(\varrho)}\left(\tau,\sigma\right)+\bar{n}_{(\theta)}^{\mu}\left(\tau\right)\Phi^{(\theta)}\left(\tau,\sigma\right).
\end{align}Once the embedding \eqref{BackStringEmbed} is specified, the equations of motion \eqref{EOMsComponent}--\eqref{ViraSoroComponent}, or equivalently the explicit solutions \eqref{BackStringTime}--\eqref{BackStringProperRadial}, determine both the tangent vectors $\partial\bar{x}^{\nu}/\partial\xi^{\mathtt{A}}$ and the orthonormal normal vectors $\bar{n}_{(\mathtt{i})}^{\mu}(\tau)$. In the conventions used above, these vectors are
\begin{align}
&\frac{\partial\bar{x}^{\mu}}{\partial\tau}\!\overset{\text{EOMs}}{=\!=\!=}\!\left(\kappa^{2}E,\varepsilon(\tau)\sqrt{\kappa^{4}E^{2}-r_{0}^{2}-\alpha_{0}^{2}\bar{\varrho}(\tau)^{2}},0,0\right),\\
&\bar{n}_{(\varrho)}^{\mu}(\tau)\!\overset{\text{EOMs}}{=\!=\!=}\!\!\left(\frac{1}{\sqrt{\kappa^{4}E^{2}\big/\left(\kappa^{4}E^{2}\!-\!r_{0}^{2}\!-\!\alpha_{0}^{2}\bar{\varrho}(\tau)^{2}\right)\!-\!1}},\frac{\varepsilon(\tau)\kappa^{2}E}{\sqrt{r_{0}^{2}+\alpha_{0}^{2}\bar{\varrho}(\tau)^{2}}},0,0\right),\\
&\frac{\partial\bar{x}^{\mu}}{\partial\sigma}=\left(0,0,0,1\right), \\
&\bar{n}_{(\theta)}^{\mu}=\left(0,0,\frac{1}{\sqrt{\alpha_{0}^{2}\bar{\varrho}(\tau)^{2}+r_{0}^{2}}},0\right).
\end{align}The quadratic action is most compactly written in terms of the extrinsic geometry of the world sheet. We therefore introduce the extrinsic curvature and the normal connection as
\begin{align}
&\Omega_{(\mathtt{i})\,\mathtt{A}\mathtt{B}}=g_{\mu\nu}\bar{n}_{(\mathtt{i})}^{\mu}\frac{\partial\bar{x}^{\alpha}}{\partial\xi^{\mathtt{A}}}\left(\partial_{\alpha}\frac{\partial\bar{x}^{\nu}}{\partial\xi^{\mathtt{B}}}+\Gamma_{\alpha\lambda}^{\nu}\frac{\partial\bar{x}^{\lambda}}{\partial\xi^{\mathtt{B}}}\right)\bigg\vert_{\partial_{r}\to\partial_{\bar{r}(\tau)}}^{r\to\bar{r}(\tau),\theta\to\frac{\pi}{2}},\\
&\mu_{(\mathtt{i})(\mathtt{j})\,\mathtt{A}}=g_{\mu\nu}\bar{n}_{(\mathtt{i})}^{\mu}\frac{\partial\bar{x}^{\alpha}}{\partial\xi^{\mathtt{A}}}\left(\partial_{\alpha}\bar{n}_{(\mathtt{j})}^{\nu}+\Gamma_{\alpha\lambda}^{\nu}\bar{n}_{(\mathtt{j})}^{\lambda}\right)\bigg\vert_{\partial_{r}\to\partial_{\bar{r}(\tau)}}^{r\to\bar{r}(\tau),\theta\to\frac{\pi}{2}},
\end{align}Here $\Omega_{(\mathtt{i})\mathtt{A}\mathtt{B}}$ is the extrinsic curvature of the world sheet, while $\mu_{(\mathtt{i})(\mathtt{j})\mathtt{A}}$ is the surface torsion, or equivalently the connection on the normal bundle. The former measures the bending of the world sheet inside the target spacetime, whereas the latter keeps track of the rotation of the transverse frame. In terms of these quantities, the covariant quadratic action for the transverse modes takes the form
\begin{align}
\nonumber
S_{(2)}&=\frac{1}{2\pi\alpha^{\prime}}\int d\tau d\sigma\sqrt{-h}\,\Phi^{(\mathtt{i})}\left\{\sum_{(\mathtt{k})}h^{\mathtt{A}\mathtt{B}}\left(\delta_{(\mathtt{i})(\mathtt{k})}\nabla_{\mathtt{A}}+\mu_{(\mathtt{i})(\mathtt{k})\mathtt{A}}\right)\right.\\
&\times\left(\delta_{(\mathtt{k})(\mathtt{j})}\nabla_{\mathtt{B}}+\mu_{(\mathtt{k})(\mathtt{j})\mathtt{B}}\right)-\mathcal{V}_{(\mathtt{i})(\mathtt{j})}\Bigg\}\Phi^{(\mathtt{j})},
\end{align}In this expression, $\nabla_{\mathtt{A}}$ denotes the covariant derivative compatible with the induced world-sheet metric $h_{\mathtt{A}\mathtt{B}}$. This form makes the covariance under world-sheet reparametrizations manifest. The matrix-valued potential $\mathcal{V}_{(\mathtt{i})(\mathtt{j})}$ is given by
\begin{align}
\label{PolarPotentialQudraticPer}
\mathcal{V}_{(\mathtt{i})(\mathtt{j})}&=h^{\mathtt{A}\mathtt{B}}R_{\mu\alpha\beta\nu}\partial_{\mathtt{A}}\bar{x}^{\mu}\partial_{\mathtt{B}}\bar{x}^{\nu}\bar{n}_{(\mathtt{i})}^{\alpha}\bar{n}_{(\mathtt{j})}^{\beta}\bigg\vert_{\partial_{r}\to\partial_{\bar{r}(\tau)}}^{r\to\bar{r}(\tau),\theta\to\frac{\pi}{2}}-\frac{2}{G_{~~\mathtt{A}_{1}}^{\mathtt{A}_{1}}}\Omega_{(\mathtt{i})\mathtt{A}\mathtt{B}}\Omega_{(\mathtt{j})}^{~\mathtt{A}\mathtt{B}},
\end{align}This potential contains the curvature-induced mass matrix of the transverse fluctuations together with the contribution from the extrinsic embedding of the world sheet. In \eqref{PolarPotentialQudraticPer}, $G_{~~\mathtt{A}_{1}}^{\mathtt{A}_{1}}$ denotes the trace of the induced metric defined in \eqref{InduceMetric}, and $R_{\mu\alpha\beta\nu}$ is the Riemann tensor of the target-space background. In the following we use the same conformal-gauge conventions as in the background analysis, namely $h^{\mathtt{A}\mathtt{B}}\to\eta^{\mathtt{A}\mathtt{B}}$ and $\nabla_{\mathtt{A}}\to\partial_{\mathtt{A}}$. This choice reduces the world-sheet covariant derivative to an ordinary derivative and leaves the quadratic fluctuation spectrum unchanged. For the wormhole metric \eqref{ProCoordiMonoWH}, the only non-vanishing components of the extrinsic curvature are
\begin{align}
\label{StringExtrinCurvatureTensor}
&\Omega_{(\varrho),\tau\tau}=\Omega_{(\varrho),\sigma\sigma}=-\frac{\varepsilon\left(\tau\right)\alpha_{0}^{2}\kappa^{2}E\bar{\varrho}\left(\tau\right)}{\sqrt{r_{0}^{2}+\alpha_{0}^{2}\bar{\varrho}\left(\tau\right)^{2}}}.
\end{align}
For this circular embedding, the normal connection is trivial: all components of $\mu_{(\mathtt{i})(\mathtt{j}),\mathtt{A}}$ vanish. Substituting the remaining geometric data into \eqref{PolarPotentialQudraticPer}, one obtains the two non-vanishing entries of the potential matrix,
\begin{align}
\label{NonVanishPotenRR}
&-\mathcal{V}_{(\varrho)(\varrho)}=\frac{\alpha_{0}^{2}\kappa^{4}E^{2}\left(2\alpha_{0}^{2}\bar{\varrho}\left(\tau\right)^{2}-r_{0}^{2}\right)}{\left(r_{0}^{2}+\alpha_{0}^{2}\bar{\varrho}(\tau)^{2}\right)^{2}}=\underbrace{\frac{\alpha_{0}^{2}\kappa^{4}E^{2}\left(-3r_{0}^{2}+2\bar{r}^{2}\right)}{\bar{r}^{4}}}_{r\text{- coordinate result}}\Bigg\vert_{r^{2}=\alpha_{0}^{2}\varrho^{2}+r_{0}^{2}},\\
\label{NonVanishPotenThetaTheta}
&-\mathcal{V}_{(\theta)(\theta)}=1+\alpha_{0}^{2}\left(\frac{\kappa^{4}E^{2}r_{0}^{2}}{\left(r_{0}^{2}+\alpha_{0}^{2}\bar{\varrho}\left(\tau\right)^{2}\right)^{2}}-1\right)=\underbrace{1-\alpha_{0}^{2}+\frac{\alpha_{0}^{2}r_{0}^{2}\kappa^{4}E^{2}}{\bar{r}^{4}}}_{r\text{- coordinate result}}\Bigg\vert_{r^{2}=\alpha_{0}^{2}\varrho^{2}+r_{0}^{2}}.
\end{align}In deriving \eqref{StringExtrinCurvatureTensor}--\eqref{NonVanishPotenThetaTheta}, we have used $\varepsilon(\tau)^2=1$. The second equalities in \eqref{NonVanishPotenRR} and \eqref{NonVanishPotenThetaTheta} reproduce the expressions obtained in normal spherical coordinates \eqref{NormalSphericalCoordinates}, as reported in \cite{Li:2026rut}. The two descriptions are related by the coordinate transformation $\varrho=\sqrt{r^{2}-r_{0}^{2}}/\alpha_{0}$, providing a useful check on the calculation. Collecting the above ingredients, the quadratic action for the two physical polarizations becomes
\begin{align}
\nonumber
S_{(2)}&=\frac{1}{2\pi\alpha^{\prime}}\int d\tau d\sigma\Bigg\{\dot{\Phi}_{\left(\varrho\right)}^{2}-\Phi_{(\varrho)}^{\prime2}+\frac{\alpha_{0}^{2}\kappa^{4}E^{2}\left(2\alpha_{0}^{2}\bar{\varrho}(\tau)^{2}-r_{0}^{2}\right)}{\left(r_{0}^{2}+\alpha_{0}^{2}\bar{\varrho}(\tau)^{2}\right)^{2}}\Phi_{(\varrho)}^{2}\\
\label{QuadraticPerString}
&+\dot{\Phi}_{(\theta)}^{2}-\Phi_{(\theta)}^{\prime2}+\left(1+\alpha_{0}^{2}\left(\frac{\kappa^{4}E^{2}r_{0}^{2}}{\left(r_{0}^{2}+\alpha_{0}^{2}\bar{\varrho}(\tau)^{2}\right)^{2}}-1\right)\right)\Phi_{(\theta)}^{2}\Bigg\}.
\end{align}Here and in what follows, an overdot and a prime denote derivatives with respect to the world-sheet coordinates $\tau$ and $\sigma$, respectively. This action is the starting point for the canonical quantization of the fluctuation modes and for the construction of the unequal-time commutators used in the OTOC analysis below.

\subsection{Canonical quantization and OTOC amplitudes}

The canonical quantization of circular strings in closely related time-dependent backgrounds was developed in Refs.~\cite{Li:2026rut,Li:2026oxy}. Applying the same framework to the quadratic action \eqref{QuadraticPerString}, we promote the two transverse fluctuations and their conjugate momenta to operators, $\Phi_{(\varrho)}, \Pi_{(\varrho)}, \Phi_{(\theta)}$ and $\Pi_{(\theta)}$. Their Fourier-mode expansions are then written as
\begin{align}
\label{QuantifyRadialPer}
&\widehat{\Phi}_{(\varrho)}\left(\tau,\sigma\right)=\frac{1}{2\pi}\sum_{n=+2}^{+\infty}\left( {\widehat{\mathrm{R}}_{n}(\tau)}\text{e}^{\text{i}n\sigma}+   {\widehat{\mathrm{R}}^{\dagger}_{n}\left(\tau\right)}\text{e}^{-\text{i}n\sigma} \right),\\
\label{QuantifyRadialMomentum}
&\hat{\Pi}_{(\varrho)}\left(\tau,\sigma\right)=\frac{1}{2\pi}\sum_{n=+2}^{+\infty}\left(\widehat{\Pi}_{n}^{(\mathrm{R})}\left(\tau\right)\text{e}^{\text{i}n\sigma}+\widehat{\Pi}_{n}^{(\mathrm{R}){\dagger}}\left(\tau\right)\text{e}^{-\text{i}n\sigma} \right),\\
\label{QuantifyAngularPer}
&\widehat{\Phi}_{(\theta)}\left(\tau,\sigma\right)=\frac{1}{2\pi}\sum_{n=+2}^{+\infty}\left(\hat{\Theta}_{n}(\tau)\text{e}^{\text{i}n\sigma}+ \hat{\Theta}_{n}^{\dagger}\left(\tau\right)\text{e}^{-\text{i}n\sigma}\right),\\
\label{QuantifyAngularMomentum}
&\hat{\Pi}_{(\theta)}\left(\tau,\sigma\right)=\frac{1}{2\pi}\sum_{n=+2}^{+\infty}\left(\widehat{\Pi}_{n}^{(\Theta)}\left(\tau\right)\text{e}^{\text{i}n\sigma}+\widehat{\Pi}_{n}^{(\Theta)\dagger}\left(\tau\right)\text{e}^{-\text{i}n\sigma}\right),
\end{align}    
where 
\begin{align}
\label{ModeExpanRadial}
\widehat{\mathrm{R}}_{n}\left(\tau\right)&= \mathcal{R}_{n}\left(\tau\right)\hat{a}_{n}^{(r)}\left(\tau_{0}\right)+\mathcal{R}_{n}^{\star}\left(\tau\right)\hat{a}_{-n}^{(r)\dagger}\left(\tau_{0}\right),\\
\label{ModeExpanRadialMomentum}
\widehat{\Pi}_{n}^{(\mathrm{R})}\left(\tau\right) &= \frac{\dot{\mathcal{R}}_{n}\left(\tau\right)}{\pi\alpha^{\prime}}\hat{a}_{n}^{(r)}\left(\tau_{0}\right)+\frac{\dot{\mathcal{R}}_{n}^{\star}\left(\tau\right)}{\pi\alpha^{\prime}}\hat{a}_{-n}^{(r)\dagger}\left(\tau_{0}\right),\\
\label{ModeExpanAngular}
\hat{\Theta}_{n}\left(\tau\right)&= \vartheta_{n}\left(\tau\right)\hat{a}_{n}^{(\theta)}\left(\tau_{0}\right)+\vartheta_{n}^{\star}\left(\tau\right)\hat{a}_{-n}^{(\theta)\dagger}\left(\tau_{0}\right),   \\
\label{ModeExpanAngularMomentum}
\widehat{\Pi}_{n}^{(\Theta)}\left(\tau\right)&= \frac{\dot{\vartheta}_{n}\left(\tau\right)}{\pi\alpha^{\prime}}\hat{a}_{n}^{(\theta)}\left(\tau_{0}\right)+\frac{\dot{\vartheta}_{n}^{\star}\left(\tau\right)}{\pi\alpha^{\prime}}\hat{a}_{-n}^{(\theta)\dagger}\left(\tau_{0}\right).
\end{align}The mode operators obey the Hermiticity condition $\hat{\mathcal{O}}_n^\dagger=\hat{\mathcal{O}}_{-n}$ for each radial and angular field/momentum mode. At the initial time, the $n$ and $-n$ oscillator pairs form two independent sets of annihilation and creation operators. Consequently, the only non-vanishing commutators are
\begin{align}
&2\pi\delta_{nm}\delta^{\mathtt{(i)}\mathtt{(j)}}\!=\left[\hat{a}_{n}^{(\mathtt{i})}\left(\tau_{0}\right),\hat{a}_{m}^{(\mathtt{j})\dagger}\left(\tau_{0}\right)\right]=\left[\hat{a}_{-n}^{(\mathtt{i})}\left(\tau_{0}\right),\hat{a}_{-m}^{(\mathtt{j})\dagger}\left(\tau_{0}\right)\right],
\end{align}where the indices $\mathtt{i}, \mathtt{j}$ label the two transverse polarizations, namely the $\varrho$ and $\theta$ directions. The notation $\tau_0$ emphasizes that the creation and annihilation operators define the initial Fock space. The subsequent time dependence is carried by the mode functions, which obey
\begin{align}
\label{GMWHGeneRadialLinearPer}
&\frac{d^{2}}{d\tau^{2}}\mathcal{R}_{n}\left(\tau\right)-\left(\frac{\alpha_{0}^{2}\kappa^{4}E^{2}\left(2\alpha_{0}^{2}\bar{\varrho}(\tau)^{2}-r_{0}^{2}\right)}{\left(r_{0}^{2}+\alpha_{0}^{2}\bar{\varrho}(\tau)^{2}\right)^{2}}-n^{2}\right)\mathcal{R}_{n}(\tau)=0,\\
\label{GMWHGeneAngularLinearPer}
&\frac{d^{2}}{d\tau^{2}}\vartheta_{n}(\tau)-\left(1-\alpha_{0}^{2}-n^{2}+\frac{\alpha_{0}^{2}\kappa^{4}E^{2}r_{0}^{2}}{\left(r_{0}^{2}+\alpha_{0}^{2}\bar{\varrho}(\tau)^{2}\right)^{2}}\right)\vartheta_{n}(\tau)=0.
\end{align}The equal-time canonical commutators of the fluctuation fields impose a normalization condition on these mode functions. Explicitly, canonical quantization requires
\begin{align}
\nonumber
\text{i}\delta\left(\sigma-\sigma^{\prime}\right)&=\left[\hat{\Phi}_{(r)}\left(\tau,\sigma\right),\hat{\Pi}_{(r)}\left(\tau,\sigma^{\prime}\right)\right]\\
&=\left[\hat{\Phi}_{(\theta)}\left(\tau,\sigma\right),\hat{\Pi}_{(\theta)}\left(\tau,\sigma^{\prime}\right)\right].
\end{align}The same requirement gives the Wronskian normalizations for the radial and angular mode functions
\begin{align}
\nonumber
\text{i}\pi\alpha^{\prime}&=\mathcal{R}_{n}\left(\tau\right)\dot{\mathcal{R}}_{n}^{\star}\left(\tau\right)-\mathcal{R}_{n}^{\star}\left(\tau\right)\dot{\mathcal{R}}_{n}\left(\tau\right)\\
\label{WronskianMode}
&=\vartheta_{n}\left(\tau\right)\dot{\vartheta}_{n}^{\star}\left(\tau\right)-\vartheta_{n}^{\star}\left(\tau\right)\dot{\vartheta}_{n}\left(\tau\right)\, , \, n\geq 2.
\end{align}To specify the initial vacuum, it is convenient to use the helicity-basis canonical variables of Ref.~\cite{Grain:2019vnq}. This linear canonical transformation treats each polarization mode as an independent oscillator. In this basis, the time-dependent annihilation operator for the radial polarization is defined by
\begin{align}
\label{LinearCanonical}
&\hat{a}_{\pm n}^{(r)}\left(\tau\right)=\frac{1}{\sqrt{2\alpha^{\prime}}}\hat{\mathrm{R}}_{\pm n}\left(\tau\right)+\text{i}\sqrt{\frac{\alpha^{\prime}}{2}}\hat{\Pi}_{\pm n}^{(\mathrm{R})}\left(\tau\right),\\
\label{LinearCanonicalAngular}
&\hat{a}_{\pm n}^{(\theta)}\left(\tau\right)=\frac{1}{\sqrt{2\alpha^{\prime}}}\hat{\Theta}_{\pm n}\left(\tau\right)+\text{i}\sqrt{\frac{\alpha^{\prime}}{2}}\hat{\Pi}_{\pm n}^{(\Theta)}\left(\tau\right),
\end{align}with the corresponding creation operator obtained by Hermitian conjugation. The angular operators $\hat{a}_{\pm n}^{(\theta)}(\tau)$ and $\hat{a}_{\mp n}^{(\theta)\dagger}(\tau)$ are introduced in the same way. These definitions fix the initial data of $\mathcal{R}_{n}(\tau)$ and $\vartheta_{n}(\tau)$ as
\begin{align}
\label{InitialConditionModes}
&\vartheta_{n}\left(\tau_{0}\right)\!=\!\frac{\text{i}}{\pi}\dot{\vartheta}_{n}\left(\tau_{0}\right)\!=\!\mathcal{R}_{n}\left(\tau_{0}\right)\!=\!\frac{\text{i}}{\pi}\dot{\mathcal{R}}_{n}\left(\tau_{0}\right)\!=\!\sqrt{\frac{\alpha^{\prime}}{2}}.
\end{align}This choice is compatible with the Wronskian condition \eqref{WronskianMode} and hence gives a well-defined vacuum at $\tau_0$. As in Refs.~\cite{Li:2026rut,Li:2026oxy}, the same choice also makes the mean particle number vanish initially. Substituting the mode expansions \eqref{ModeExpanRadial}-\eqref{ModeExpanAngularMomentum} into the quadratic action \eqref{QuadraticPerString}, one obtains the Hamiltonian operators
\begin{align}
\nonumber
\hat{\mathcal{H}}_{(\varrho)}^{\text{(2)}}&=\frac{1}{4\pi^{2}}\sum_{n=2}^{\infty}\bigg\{\left(n^{2}-\frac{\alpha_{0}^{2}\kappa^{4}E^{2}(2\alpha_{0}^{2}\bar{\varrho}^{2}-r_{0}^{2})}{(r_{0}^{2}+\alpha_{0}^{2}\bar{\varrho}^{2})^{2}}+\pi^{2}\right)(\hat{a}_{n}^{(\varrho)\dagger}\hat{a}_{n}^{(\varrho)}+\hat{a}_{-n}^{(\varrho)}\hat{a}_{-n}^{(\varrho)\dagger})\\
\label{HamilQuadraticPerInr}
&+\left(n^{2}-\frac{\alpha_{0}^{2}\kappa^{4}E^{2}(2\alpha_{0}^{2}\bar{\varrho}^{2}-r_{0}^{2})}{(r_{0}^{2}+\alpha_{0}^{2}\bar{\varrho}^{2})^{2}}-\pi^{2}\right)(\hat{a}_{n}^{(\varrho)\dagger}\hat{a}_{-n}^{(\varrho)\dagger}+\hat{a}_{-n}^{(\varrho)}\hat{a}_{n}^{(\varrho)})\bigg\},\\
\nonumber
\hat{\mathcal{H}}_{(\theta)}^{\text{(2)}}&=\frac{1}{4\pi^{2}}\sum_{n=2}^{\infty}\bigg\{\left(n^{2}-1+\alpha_{0}^{2}-\frac{\kappa^{4}\alpha_{0}^{2}E^{2}r_{0}^{2}}{(r_{0}^{2}+\alpha_{0}^{2}\bar{\varrho}^{2})^{2}}+\pi^{2}\right)(\hat{a}_{n}^{(\theta)\dagger}\hat{a}_{n}^{(\theta)}+\hat{a}_{-n}^{(\theta)}\hat{a}_{-n}^{(\theta)\dagger})\\
\label{HamilQuadraticPerInTheta}
&+\left(n^{2}-1+\alpha_{0}^{2}-\frac{\kappa^{4}\alpha_{0}^{2}E^{2}r_{0}^{2}}{(r_{0}^{2}+\alpha_{0}^{2}\bar{\varrho}^{2})^{2}}-\pi^{2}\right)(\hat{a}_{n}^{(\theta)\dagger}\hat{a}_{-n}^{(\theta)\dagger}+\hat{a}_{-n}^{(\theta)}\hat{a}_{n}^{(\theta)})\bigg\}.
\end{align}Starting from the relations \eqref{LinearCanonical}-\eqref{LinearCanonicalAngular} directly or from the symmetry structure built on the Hamiltonian operators \eqref{HamilQuadraticPerInr}-\eqref{HamilQuadraticPerInTheta}, the particle number can be evaluated as \cite{Li:2026rut}
\begin{align}
\nonumber
\mathcal{N}_{n}^{(\mathtt{i})}\left(\tau\right)&={}_{\tau_{0}}\langle\tilde{0}_{n},\tilde{0}_{-n}\vert\frac{1}{2\pi}\sum_{m=2}^{\infty}\hat{a}_{m}^{(\mathtt{i})\dagger}\left(\tau\right)\hat{a}_{n}^{(\mathtt{i})}\left(\tau\right)\vert\tilde{0}_{n},\tilde{0}_{-n}\rangle_{\tau_{0}}\\
\label{ParticleNum}
&=\begin{cases}
\begin{array}{c}
\frac{1}{2\alpha^{\prime}}\bigg\vert\mathcal{R}_{n}(\tau)-\frac{\text{i}}{\pi}\dot{\mathcal{R}}_{n}(\tau)\bigg\vert^{2}\,,\,\mathtt{i}=\varrho\\
\frac{1}{2\alpha^{\prime}}\bigg\vert\vartheta_{n}(\tau)-\frac{\text{i}}{\pi}\dot{\vartheta}_{n}(\tau)\bigg\vert^{2}\,,\,\mathtt{i}=\theta
\end{array}\end{cases},
\end{align}where $\tilde{m}$ and $\tilde{n}$ denote occupation numbers of the quantum state. We use tildes to distinguish them from the Fourier mode labels $m$ and $n$ appearing in \eqref{QuantifyRadialPer}--\eqref{QuantifyAngularMomentum}. Equation \eqref{ParticleNum} is consistent with \eqref{InitialConditionModes}: at $\tau=\tau_0$, the particle number vanishes.
 
With the quantized fluctuation modes in hand, we now construct the out-of-time-ordered correlator (OTOC) as a diagnostic of quantum chaos. OTOCs have become a standard probe of operator growth and information scrambling in holographic black holes and shock-wave geometries \cite{Shenker:2013pqa,Roberts:2014Localized,Maldacena:2015waa}, as well as in nearly conformal/SYK-like systems \cite{Maldacena:2016hyu}. More specifically, in gravitational systems they have been used to characterize near-horizon butterfly dynamics, localized shocks and stringy corrections to black-hole scrambling \cite{Roberts:2014Localized,Shenker:2014Stringy}, the Schwarzian dynamics of nearly AdS$_2$ gravity \cite{Maldacena:2016NAdS2}, maximal chaos in open-string and D-brane probes \cite{Banerjee:2018Strings}, and cosmological scrambling in de Sitter or squeezed-state settings \cite{Aalsma:2020DeSitter,Haque:2020SqueezedOTOC}. They are also widely used in quantum many-body lattice dynamics \cite{Luitz:2017jvm}. These examples motivate the present use of OTOCs as a local probe of whether the time-dependent throat potential of a traversable wormhole can generate quantum chaotic growth in the transverse string sector. In the present probe-string system, the OTOC amplitudes for each Fourier mode are extracted from the corresponding double commutators,
\begin{small}
\begin{align}
\nonumber
(2\pi)^{2}\delta_{nn^{\prime}}\delta_{mm^{\prime}}\mathcal{C}_{n,m}^{(\hat{\mathcal{O}}_{1};\hat{\mathcal{O}}_{2})}&=\!-\left[\hat{\mathcal{O}}_{1,n}(\tau),\hat{\mathcal{O}}_{2,-n^{\prime}}(\tau_{0})\right]
\times\left[\hat{\mathcal{O}}_{1,m}(\tau),\hat{\mathcal{O}}_{2,-m^{\prime}}(\tau_{0})\right].
\end{align}
\end{small}In principle, one can form four OTOC amplitudes from the fluctuation and momentum operators of a given polarization. For the radial sector, the two elementary operators are $\hat{\mathcal{O}}_{1,2}=\hat{\mathrm{R}},\hat{\Pi}^{(\mathrm{R})}$, with an analogous pair in the angular sector. For the Lyapunov analysis below we focus on the position-momentum commutator,
\begin{align}
\nonumber
\mathcal{C}_{n,m}^{(\hat{\mathrm{R}};\widehat{\Pi}^{(\mathrm{R})})}(\tau,\tilde{\tau})&=-\frac{1}{(2\pi)^{2}}\sum_{n^{\prime}}\sum_{m^{\prime}}\left[\hat{\mathrm{R}}_{n}(\tau),\hat{\Pi}_{-n^{\prime}}^{(\mathrm{R})}(\tilde{\tau})\right]\left[\hat{\mathrm{R}}_{m}(\tau),\hat{\Pi}_{-m^{\prime}}^{(\mathrm{R})}(\tilde{\tau})\right]\\
&=4\,\text{Im}\left(\frac{\mathcal{R}_{n}(\tau)\dot{\mathcal{R}}_{n}^{\star}(\tilde{\tau})}{\pi\alpha^{\prime}}\right)\,\text{Im}\left(\frac{\mathcal{R}_{m}(\tau)\dot{\mathcal{R}}_{m}^{\star}(\tilde{\tau})}{\pi\alpha^{\prime}}\right).
\end{align}
This is the direct field-theoretic analogue of the standard quantum-mechanical OTOC,
\begin{align}
\label{StandardOTOCInQM}
&\mathcal{C}_{T}(t,t_0)=-\left\langle \left[\hat{x}(t),\hat{p}(t_0)\right]^2 \right\rangle,
\end{align}
as discussed in Refs.~\cite{Hashimoto:2017oit,Xu:2019lhc}. Following the convention \eqref{StandardOTOCInQM}, the corresponding real-space unequal-time commutator is obtained by summing over the Fourier modes,
\begin{align}
\nonumber
\mathcal{C}_{T}^{(\varrho)}(\tau,\tilde{\tau})&=-\left\langle\left[\widehat{\Phi}_{(\varrho)}\left(\tau,\sigma\right),\hat{\Pi}_{(\varrho)}\left(\tilde{\tau},\sigma\right)\right]^{2}\right\rangle=\frac{1}{\pi^{2}}\sum_{n=+2}^{+\infty}\sum_{m=+2}^{+\infty}\mathcal{C}_{n,m}^{(\hat{\mathrm{R}};\widehat{\Pi}^{(\mathrm{R})})}\left(\tau,\tilde{\tau}\right) \\
&=\frac{2}{\pi}\sum_{n=+2}^{+\infty}\text{Im}\left(\frac{\mathcal{R}_{n}(\tau)\dot{\mathcal{R}}_{n}^{\star}\left(\tilde{\tau}\right)}{\pi\alpha^{\prime}}\right)\times\frac{2}{\pi}\sum_{m=+2}^{+\infty}\text{Im}\left(\frac{\mathcal{R}_{m}(\tau)\dot{\mathcal{R}}_{m}^{\star}\left(\tilde{\tau}\right)}{\pi\alpha^{\prime}}\right).
\end{align}Thus the radial OTOC amplitude used in the numerical analysis can be written as
\begin{align}
\label{FinalOTOCamplitudeRadial}
&\sqrt{\mathcal{C}^{(\varrho)}_{T}\left(\tau,\tilde{\tau}\right)}=\frac{2}{\alpha^{\prime}\pi^{2}}\bigg\vert\sum_{n=+2}^{+\infty}\text{Im}\left(\mathcal{R}_{n}(\tau)\dot{\mathcal{R}}_{n}^{\star}\left(\tilde{\tau}\right)\right)\bigg\vert.
\end{align}The angular polarization is treated in the same way, giving
\begin{align}
\label{FinalOTOCamplitudeAngular}
&\sqrt{\mathcal{C}_{T}^{(\theta)}\left(\tau,\tilde{\tau}\right)}=\frac{2}{\alpha^{\prime}\pi^{2}}\Bigg\vert\sum_{n=+2}^{+\infty}\text{Im}\left(\vartheta_{n}(\tau)\dot{\vartheta}_{n}^{\star}(\tilde{\tau})\right)\Bigg\vert.
\end{align}

This subsection therefore converts the quantized fluctuation problem into the OTOC observables that will be used below. The essential output is the pair of unequal-time commutator amplitudes \eqref{FinalOTOCamplitudeRadial} and \eqref{FinalOTOCamplitudeAngular}, whose time dependence is entirely determined by the radial and angular mode functions. This construction isolates the main probe-sector mechanism of the paper: possible Lyapunov growth is not inferred from the classical circular trajectory itself, but from quantum transverse fluctuations driven by the time-dependent wormhole geometry. In the next section we solve these mode equations numerically and extract the corresponding effective Lyapunov exponents.

\section{Numerical analysis about the Lyapunov exponent \label{NumAnaOTOC}}

\subsection{Ellis-Bronnikov wormhole case \label{PureEBWHCase}}

We first consider the Ellis-Bronnikov wormhole, which can be obtained from the geometry \eqref{NormalSphericalCoordinates} by taking an appropriate limit. In particular, starting from the metric \eqref{ProCoordiMonoWH}, this limit is implemented by sending $\eta\to0$ while keeping the throat radius finite through $L_0\propto1/\eta$, so that $\alpha_0\to1$. In this limit the throat radius is the only remaining geometric scale. Since the Ricci curvature at the throat behaves as $R_{\text{Ricci}}\sim-2/r_0^2$, a string crossing the throat probes an effective tidal field set by the probe energy relative to the throat scale. We therefore introduce the dimensionless parameter $\chi=\kappa^{2}E/r_0$, which characterizes the tidal strength felt during the throat crossing. Although a traversable wormhole has no horizon and hence no genuine surface gravity, $\chi$ plays an analogous role for the probe dynamics: it measures how strongly the throat drives the time-dependent transverse fluctuation potentials. The first goal of this section is to test whether quantum chaos is generated in the vicinity of the wormhole throat. Once such growth is identified, it is natural to organize the numerical data in terms of the dimensionless combination $\chi=\kappa^{2}E/r_0$ and to study its relation to the effective Lyapunov exponent $\lambda_\tau$ extracted from the OTOC amplitudes.

We begin by plotting the radial and angular amplitudes $\sqrt{\mathcal{C}^{(\varrho)}_{T}(\tau,\tilde{\tau})}$ and $\sqrt{\mathcal{C}_{T}^{(\theta)}(\tau,\tilde{\tau})}$, defined in \eqref{FinalOTOCamplitudeRadial}--\eqref{FinalOTOCamplitudeAngular}, as functions of the world-sheet time $\tau$. In principle, the sums over the Fourier or winding number $n$ extend to infinity. The mode equations, however, show that only a finite range of modes can contribute efficiently to the growth near the throat. More explicitly, in the Ellis-Bronnikov wormhole the perturbation equations \eqref{GMWHGeneRadialLinearPer}-\eqref{GMWHGeneAngularLinearPer} reduce to 
\begin{align}
\label{EffOmegaSquareRadialEBWH}
&\ddot{\mathcal{R}}_{n}\left(\tau\right)=\omega_{n}^{(\mathcal{R})}\left(\tau\right)^{2}\,\mathcal{R}_{n}\left(\tau\right)~,~\omega_{n}^{(\mathcal{R})}\left(\tau\right)^{2}=\frac{\left(2(\chi^{2}-1)\cos^{2}\left(\tau-\tau_{0}\right)-1\right)}{\left((\chi^{2}-1)\cos^{2}\left(\tau-\tau_{0}\right)+1\right)^{2}}\chi^{2}-n^{2},\\
\label{EffOmegaSquareAngularEBWH}
&\ddot{\vartheta}_{n}\left(\tau\right)=\omega_{n}^{(\vartheta)}\left(\tau\right)^{2}\,\vartheta_{n}(\tau)~,~\omega_{n}^{(\vartheta)}\left(\tau\right)^{2}=\frac{1}{\left(\left(\chi^{2}-1\right)\cos^{2}\left(\tau-\tau_{0}\right)+1\right)^{2}}\chi^{2}-n^{2}.
\end{align}These equations make the mode truncation transparent. The integer mode number contributes the universal term $-n^2$, while the throat generates a bounded time-dependent potential whose overall size is controlled by $\chi$. At the first throat crossing, where $\bar{\varrho}=0$, the angular channel behaves as $\omega_n^{(\vartheta)}(\tau_{\text{throat}})^2\sim\chi^2-n^2$. Therefore only a finite band of angular modes with $n\lesssim\chi$ can be locally amplified, whereas sufficiently large $n$ gives oscillatory solutions. By contrast, the radial channel gives $\omega_n^{(\mathcal{R})}(\tau_{\text{throat}})^2\sim-(n^2+\chi^2)$, and is already oscillatory at the exact crossing. Any radial enhancement must therefore come from the finite-time profile of the potential along the trajectory rather than from an instability of arbitrarily high modes. 

\begin{figure}
 \begin{center}
    \includegraphics[scale=0.55]{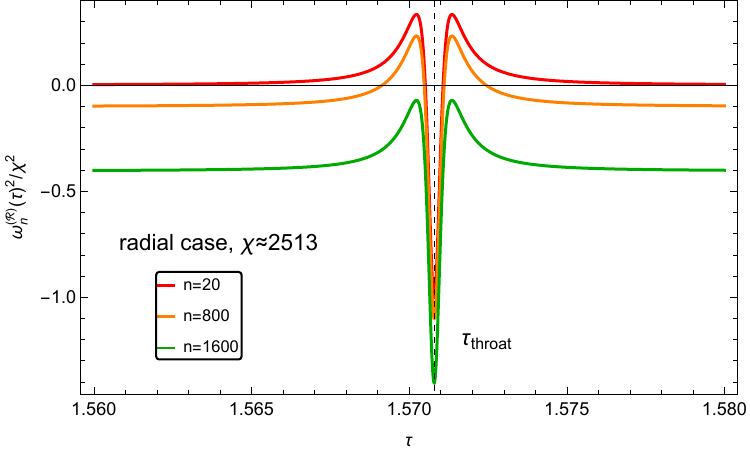}
    \includegraphics[scale=0.55]{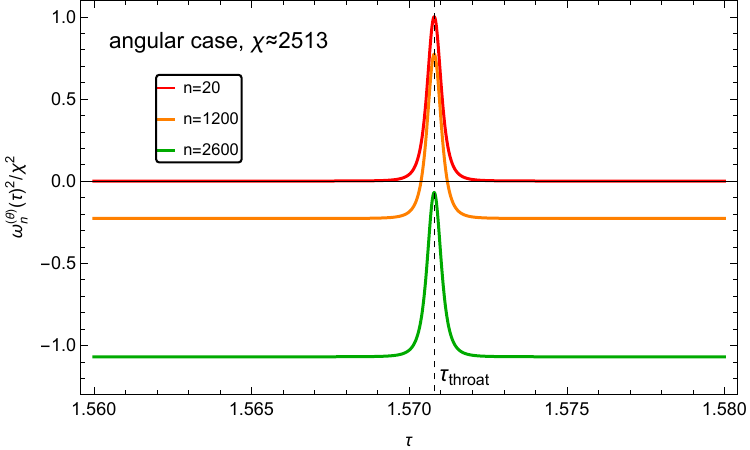}\\
    \includegraphics[scale=0.55]{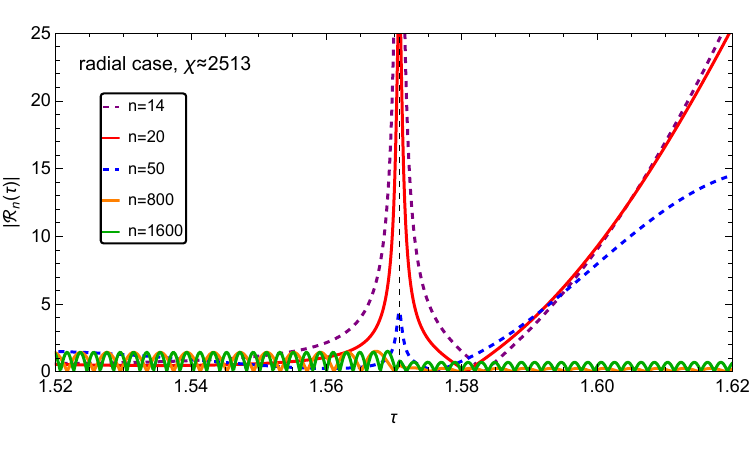}
    \includegraphics[scale=0.55]{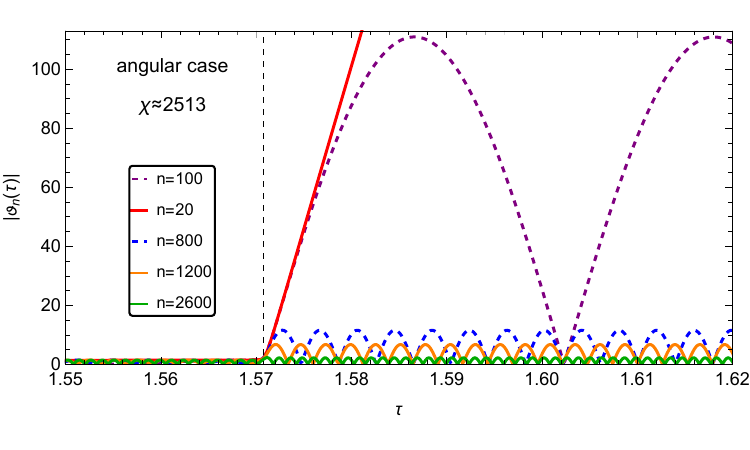}
 \caption{Effective frequency profiles and the corresponding mode-function response for the Ellis-Bronnikov wormhole. The upper panels show the time dependence of $\omega_{n}^{(\mathcal{R})}(\tau)^{2}$ and $\omega_{n}^{(\vartheta)}(\tau)^{2}$ for representative values of the world-sheet mode number $n$, with particular emphasis on the neighborhood of the wormhole throat. The lower panels show the corresponding evolution of the mode-function moduli $\vert\mathcal{R}_{n}(\tau)\vert$ and $\vert\vartheta_{n}(\tau)\vert$. For convenience and without loss of generality, we set $\tau_0=0$. The parameters are chosen as $G=1$, $M_{\text{pl}}=1/\sqrt{G}=1$, $\kappa=\sqrt{8\pi G}$, $E=1000\,M_{\text{pl}}$ and $r_0=10\,G M_{\text{pl}}$, giving $\chi=\kappa^{2}E/r_0\simeq2513$.}
 \label{EBWHEffOmegaSquare}
 \end{center}
 \end{figure}

To display the competition between the $-n^2$ term and the time-dependent throat potential, we plot $\omega_{n}^{(\mathcal{R})}(\tau)^{2}$ and $\omega_{n}^{(\vartheta)}(\tau)^{2}$ near the throat in the upper panels of Fig.~\ref{EBWHEffOmegaSquare}. Since the potentials in \eqref{EffOmegaSquareRadialEBWH}--\eqref{EffOmegaSquareAngularEBWH} depend on $\cos^2(\tau-\tau_0)$, it is sufficient to follow one period, for instance $\tau_0\leq\tau\leq\tau_0+\pi$. Moreover, the positive part of the time-dependent contribution reaches its maximum in the vicinity of the throat. The plotted window therefore captures the region where the competition with $n^2$ is most relevant.
The behavior displayed in Fig.~\ref{EBWHEffOmegaSquare} illustrates this mechanism. As $n$ increases, the contribution from the throat potential becomes subleading compared with the $n^2$ term, and the corresponding mode functions rapidly approach an oscillatory regime. More generally, when $n^2$ is much larger than the maximal size of the time-dependent potential, both polarizations are well described by adiabatic WKB oscillations. The associated non-adiabatic mixing, and hence the contribution to the growth of the OTOC amplitudes, is then suppressed. This provides a controlled criterion for truncating the mode sum in the numerical evaluation. 
 
\begin{figure}
 \begin{center}
    \includegraphics[scale=0.55]{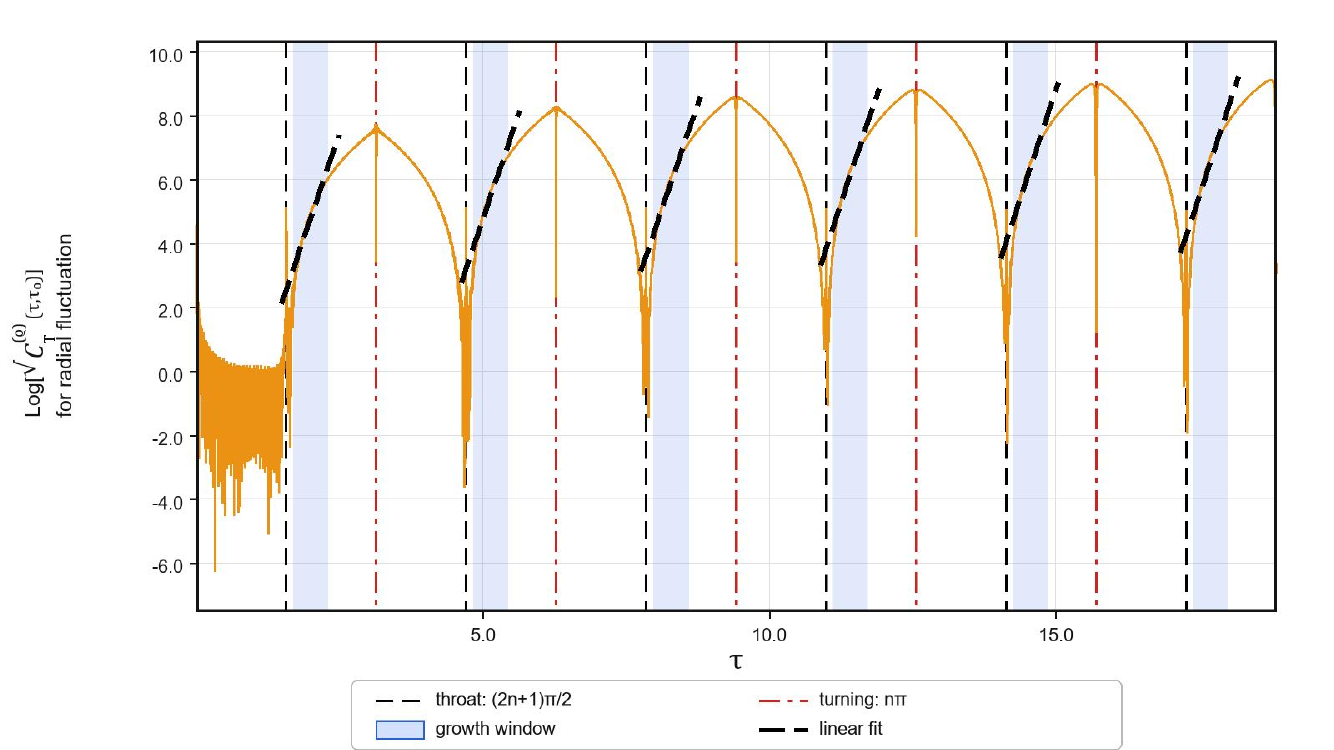}\\
    \includegraphics[scale=0.55]{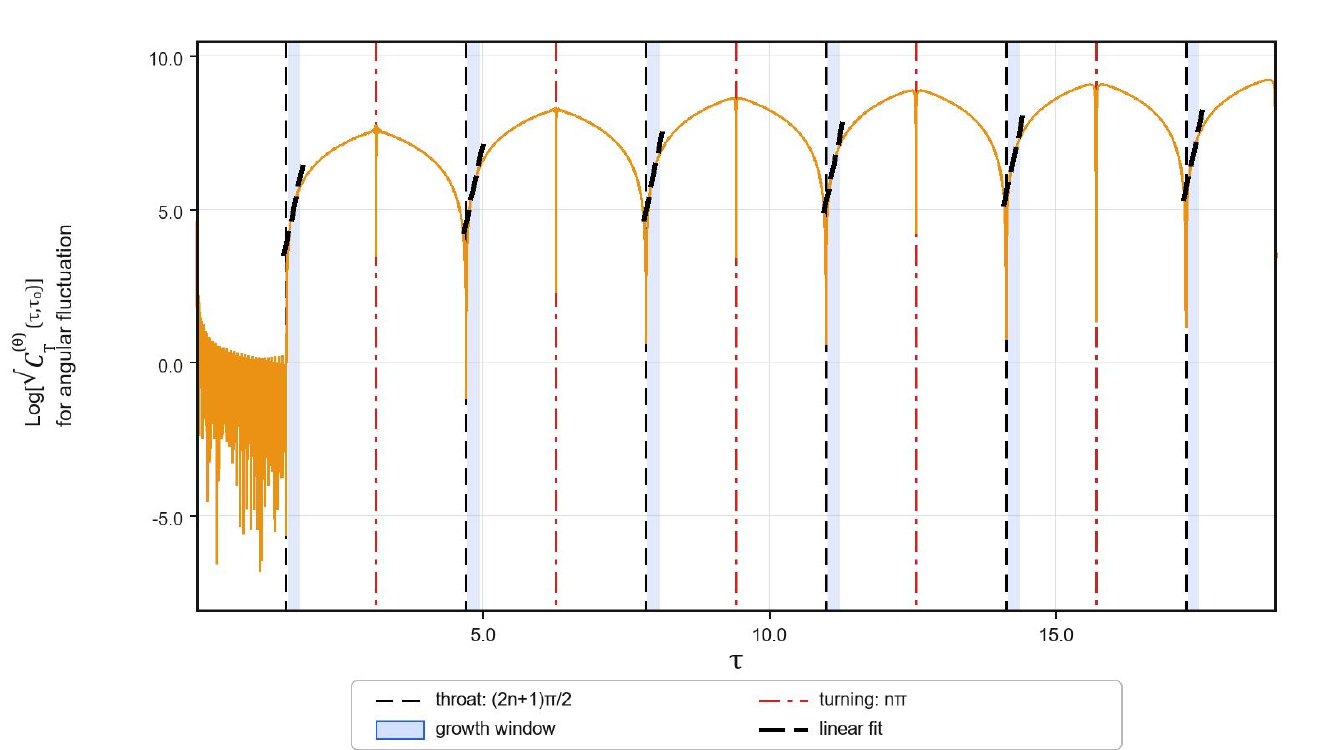}
 \caption{Representative OTOC amplitudes for the Ellis-Bronnikov wormhole, computed from \eqref{FinalOTOCamplitudeRadial} and \eqref{FinalOTOCamplitudeAngular}. The numerical parameters are the same as those used in Fig.~\ref{EBWHEffOmegaSquare}. The shaded intervals indicate the growth windows used to fit the logarithm of the unequal-time commutator, and the short black dashed segments show the corresponding linear fits. The black vertical dashed lines mark the world-sheet times $\tau_{\text{throat}}=(2k+1)\pi/2$ at which the probe string reaches the wormhole throat, whereas the red vertical dash-dotted lines mark the times $\tau_{\text{turning}}=k\pi$ at which its radial velocity vanishes at the distant turning points, with $k\in\mathbb{Z}$.}
 \label{CTvsTauAndLinearGrowth}
 \end{center}
 \end{figure}

 Having established a controlled truncation of the mode sum, we next determine the OTOC evolution numerically. For a fixed value of $\chi$, we solve the radial and angular mode equations \eqref{EffOmegaSquareRadialEBWH}--\eqref{EffOmegaSquareAngularEBWH}, subject to the initial conditions specified in \eqref{InitialConditionModes}. The resulting numerical mode functions are then substituted into the radial and angular OTOC amplitudes \eqref{FinalOTOCamplitudeRadial} and \eqref{FinalOTOCamplitudeAngular}, respectively, with the reference time fixed at $\tilde{\tau}=\tau_0$. Their time evolution is displayed in Fig.~\ref{CTvsTauAndLinearGrowth}. We plot the logarithm of each OTOC amplitude, $\mathcal{A}_{\mathtt{i}}(\tau)\equiv\sqrt{\mathcal{C}_{T}^{(\mathtt{i})}(\tau,\tau_0)}$, because a transient exponential-growth law $\mathcal{A}_{\mathtt{i}}(\tau)\simeq\mathcal{A}_{\mathtt{i},0}\exp[\lambda_{\tau}^{(\mathtt{i})}(\tau-\tau_{\mathrm{a}})]$ becomes linear on this scale. We identify the finite interval immediately following the first throat passage in which $\log\mathcal{A}_{\mathtt{i}}$ exhibits approximately linear growth and perform a least-squares linear regression within that interval. The fitted slope defines the effective finite-time Lyapunov exponent $\lambda_{\tau}^{(\mathtt{i})}$. The qualifier ``effective'' emphasizes that the background is time dependent and that the exponent is extracted from a finite growth window rather than from an asymptotic late-time limit. The shaded regions in Fig.~\ref{CTvsTauAndLinearGrowth} indicate the selected fitting windows, while the short black dashed segments show the corresponding linear fits.

\begin{figure}
 \begin{center}
    \includegraphics[scale=0.55]{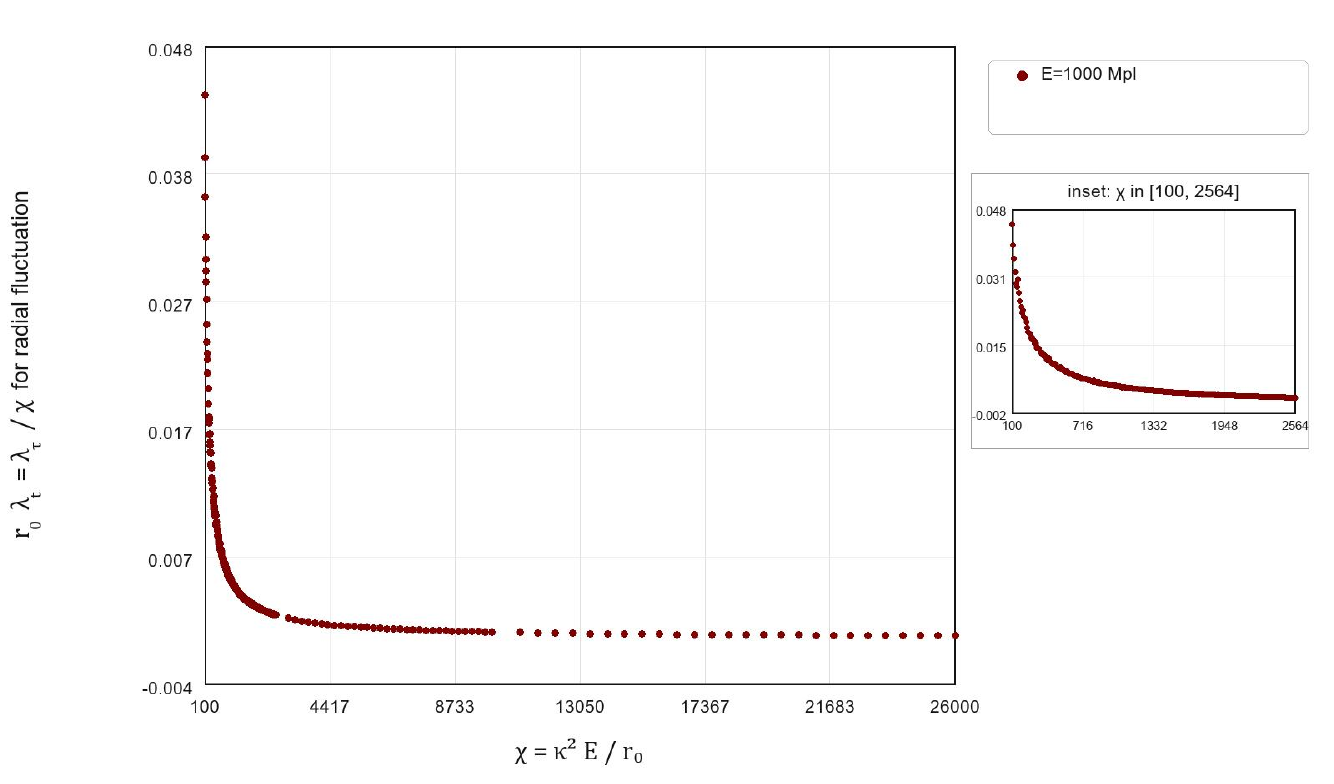}\\
    \includegraphics[scale=0.55]{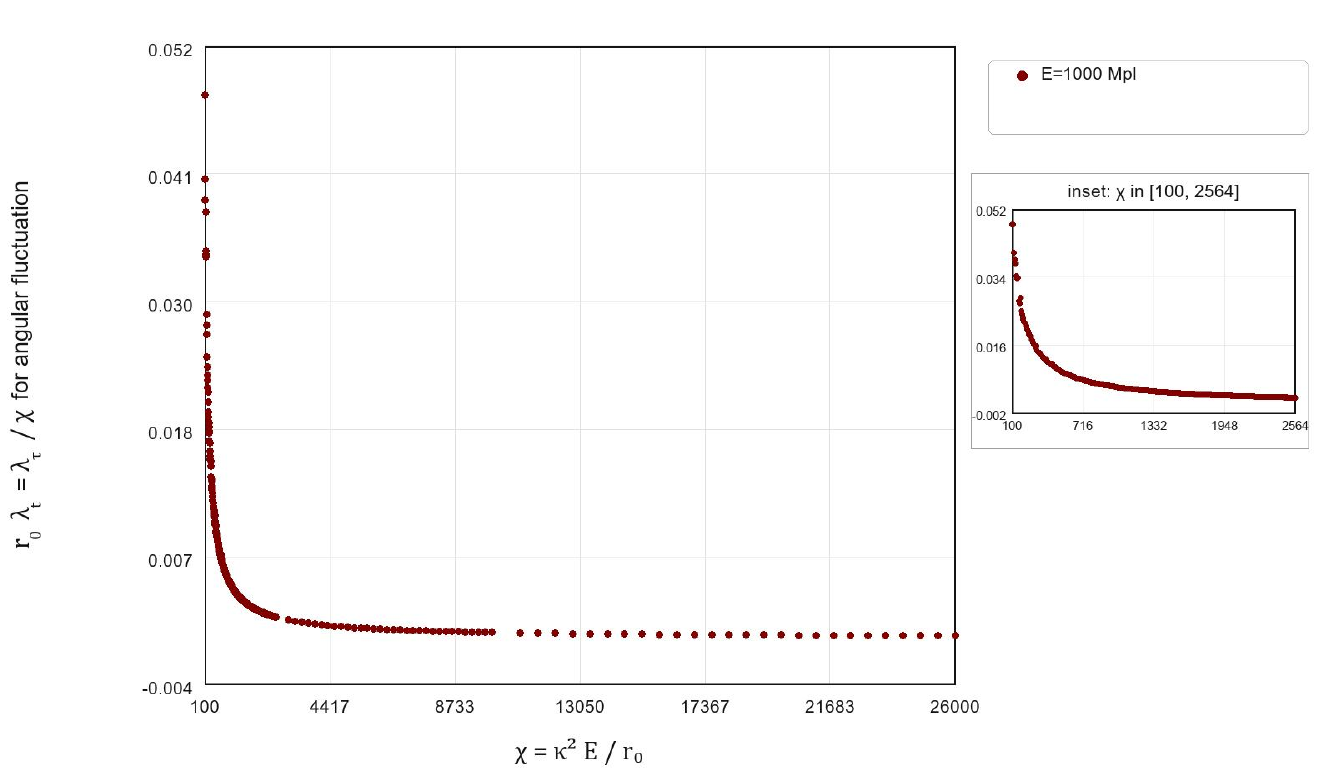}
 \caption{Dimensionless physical-time Lyapunov rate $r_0\lambda_t$ as a function of the throat parameter $\chi=\kappa^{2}E/r_0$ for the Ellis-Bronnikov wormhole. The upper and lower panels show the radial and angular fluctuation channels, respectively. For each value of $\chi$, the OTOC amplitude is evolved as in Fig.~\ref{CTvsTauAndLinearGrowth}, and the world-sheet Lyapunov exponent $\lambda_\tau$ is extracted from the first growth window associated with the initial passage through the throat. Since $t=\kappa^{2}E\tau+t_0$, the Lyapunov exponent defined with respect to the physical time is $\lambda_t=\lambda_\tau/(\kappa^{2}E)$. In natural units $\lambda_t$ has dimensions of inverse time, or equivalently energy, so multiplication by $r_0$ gives the dimensionless combination $r_0\lambda_t=\lambda_\tau/\chi$. This quantity measures the Lyapunov growth accumulated by the circular string over a characteristic throat timescale. In evaluating the OTOC amplitude at each $\chi$, the upper limit of the world-sheet mode-number sum, $n_{\mathrm{cutoff}}$, is chosen to be the integer nearest to $\chi$.}
 \label{LambdaVsR0ForRadial}
 \end{center}
 \end{figure}

For any prescribed value of $\chi$, the procedure described above yields the corresponding effective Lyapunov exponent. Repeating the calculation across the parameter range allows us to study the dimensionless physical-time rate $r_0\lambda_t$ as a function of $\chi$, as shown in Fig.~\ref{LambdaVsR0ForRadial}. Here $\lambda_t=\lambda_\tau/(\kappa^{2}E)$ is the Lyapunov exponent defined with respect to the physical time $t$, so that $r_0\lambda_t=\lambda_\tau/\chi$ measures the number of Lyapunov e-foldings accumulated over a characteristic throat timescale. Figure~\ref{CTvsTauAndLinearGrowth} shows that both polarizations develop intervals in which the logarithm of the OTOC amplitude grows approximately linearly after the probe crosses the throat. The synchronization of these growth windows with the throat passages indicates that the enhancement is driven by the localized, time-dependent tidal potential rather than by the distant turning points of the classical trajectory. The different radial and angular growth rates further show that the response is polarization dependent, reflecting the distinct effective potentials in \eqref{EffOmegaSquareRadialEBWH} and \eqref{EffOmegaSquareAngularEBWH}. After conversion to physical time and normalization by $r_0$, Fig.~\ref{LambdaVsR0ForRadial} shows that $r_0\lambda_t$ is positive in both channels over the displayed range and decreases as $\chi$ is increased. Thus, although increasing $\chi$ sharpens the instantaneous throat-induced tidal driving, the net Lyapunov growth accumulated within one characteristic throat timescale becomes weaker at large $\chi$. This behavior characterizes a nontrivial dynamical crossover in the probe sector and should not, by itself, be interpreted as a thermodynamic phase transition of the wormhole geometry. Most importantly, when combined with the absence of a classical Lyapunov instability for the background circular trajectory, these numerical results provide direct evidence, within the OTOC diagnostic, that quantum transverse fluctuations of the probe string can develop a throat-induced quantum-chaotic response even when the underlying classical motion remains stable.

\subsection{Effects of the deficit angle on the Lyapunov exponent \label{TopoGMWHCase}}

The central result of this work was already established in Sec.~\ref{PureEBWHCase}: in the canonical Ellis--Bronnikov background, quantum fluctuations of a circular probe string develop a finite-time exponentially growing OTOC signal during a throat passage. We now address a distinct question, namely whether a topological defect in the wormhole geometry enhances or suppresses this throat-induced quantum-chaotic response. This question is motivated by our previous analysis in Ref.~\cite{Li:2026rut}, where the quantized circular-string fluctuations were described in terms of two-mode squeezed states and their particle--antiparticle entanglement. For the global-monopole wormhole branch, the entanglement generated by a throat passage was found to increase with the deficit factor $\alpha_0$. Since a larger solid-angle deficit corresponds to a smaller $\alpha_0$, this result means that a stronger defect suppresses the quantum correlations carried by the probe-string fluctuation sector. Entanglement and OTOC growth diagnose different aspects of the quantum dynamics, so this observation does not by itself determine the scrambling behavior. It nevertheless suggests a natural hypothesis: increasing the deficit may weaken the throat-induced amplification of the fluctuations and hence reduce the effective Lyapunov exponent, possibly quenching the chaotic signal once the defect becomes sufficiently strong.

To test this hypothesis directly, we retain a finite deficit angle in the fluctuation equations. The defect modifies both the strength of the time-dependent potentials, through factors of $\alpha_0^2$, and their world-sheet time dependence, through the phase $\alpha_0(\tau-\tau_0)$. The radial and angular mode equations then become
\begin{align}
\label{EffOmegaSquareRadialGMWH}
&\ddot{\mathcal{R}}_{n}(\tau)=\omega_{n}^{(\mathcal{R})}(\tau)^{2}\mathcal{R}_{n}(\tau)~,~\omega_{n}^{(\mathcal{R})}(\tau)^{2}=\frac{2(\chi^{2}-1)\cos^{2}\left(\alpha_{0}(\tau-\tau_{0})\right)-1}{\left(\left(\chi^{2}-1\right)\cos^{2}\left(\alpha_{0}\left(\tau-\tau_{0}\right)\right)+1\right)^{2}}\chi^{2}\alpha_{0}^{2}-n^{2},\\
\label{EffOmegaSquareAngularGMWH}
&\ddot{\vartheta}_{n}(\tau)\!=\!\omega_{n}^{(\vartheta)}(\tau)^{2}\vartheta_{n}(\tau)\,,\,\omega_{n}^{(\vartheta)}(\tau)^{2}\!=\!1\!-\!\alpha_{0}^{2}\!+\frac{1}{\left((\chi^{2}\!-\!1)\cos^{2}\left(\alpha_{0}(\tau-\tau_{0})\right)\!+\!1\right)^{2}}\chi^{2}\alpha_{0}^{2}\!-\!n^{2}.
\end{align}At the throat, the angular effective frequency in \eqref{EffOmegaSquareAngularGMWH} reduces to $\omega_{n}^{(\vartheta)}(\tau_{\text{throat}})^{2}=1+(\chi^{2}-1)\alpha_{0}^{2}-n^{2}$. Local exponential amplification is therefore possible only for $n^{2}<1+(\chi^{2}-1)\alpha_{0}^{2}$, which becomes $n\lesssim\alpha_0\chi$ in the regime $\chi\gg1$. Equivalently, for a fixed mode with $n>1$, the angular throat instability closes when $\alpha_0\leq\sqrt{(n^{2}-1)/(\chi^{2}-1)}$. The radial channel is qualitatively different: Eq.~\eqref{EffOmegaSquareRadialGMWH} gives $\omega_{n}^{(\mathcal{R})}(\tau_{\text{throat}})^{2}=-\alpha_0^{2}\chi^{2}-n^{2}<0$, so its possible amplification is controlled by the finite off-throat intervals in which the complete time-dependent profile becomes positive, rather than by the value at the throat alone. For each pair $(\chi,\alpha_0)$, we solve Eqs.~\eqref{EffOmegaSquareRadialGMWH} and \eqref{EffOmegaSquareAngularGMWH} numerically using the initial conditions in Eq.~\eqref{InitialConditionModes}, and substitute the resulting mode functions into the radial and angular OTOC amplitudes \eqref{FinalOTOCamplitudeRadial} and \eqref{FinalOTOCamplitudeAngular}. Applying the same first-throat-crossing fit and conversion to $r_0\lambda_t$ as in Sec.~\ref{PureEBWHCase} gives the results in Fig.~\ref{LambdaVsR0ForRadial}. The radial channel displays a clear reduction of the extracted rate as $\alpha_0$ decreases, with a sufficiently strong deficit driving it close to zero. The angular channel exhibits the same overall suppression at moderate and large $\chi$, although its low-$\chi$ ordering is locally non-monotonic. This residual structure shows that the defect acts through the polarization-dependent fluctuation potentials rather than through a universal rescaling of the Lyapunov exponent. Nevertheless, the strong-defect behavior supports the hypothesis suggested by Ref.~\cite{Li:2026rut}: the suppression of probe-sector quantum correlations is accompanied by a suppression of OTOC growth.

\begin{figure}[H]
 \begin{center}
    \includegraphics[scale=0.55]{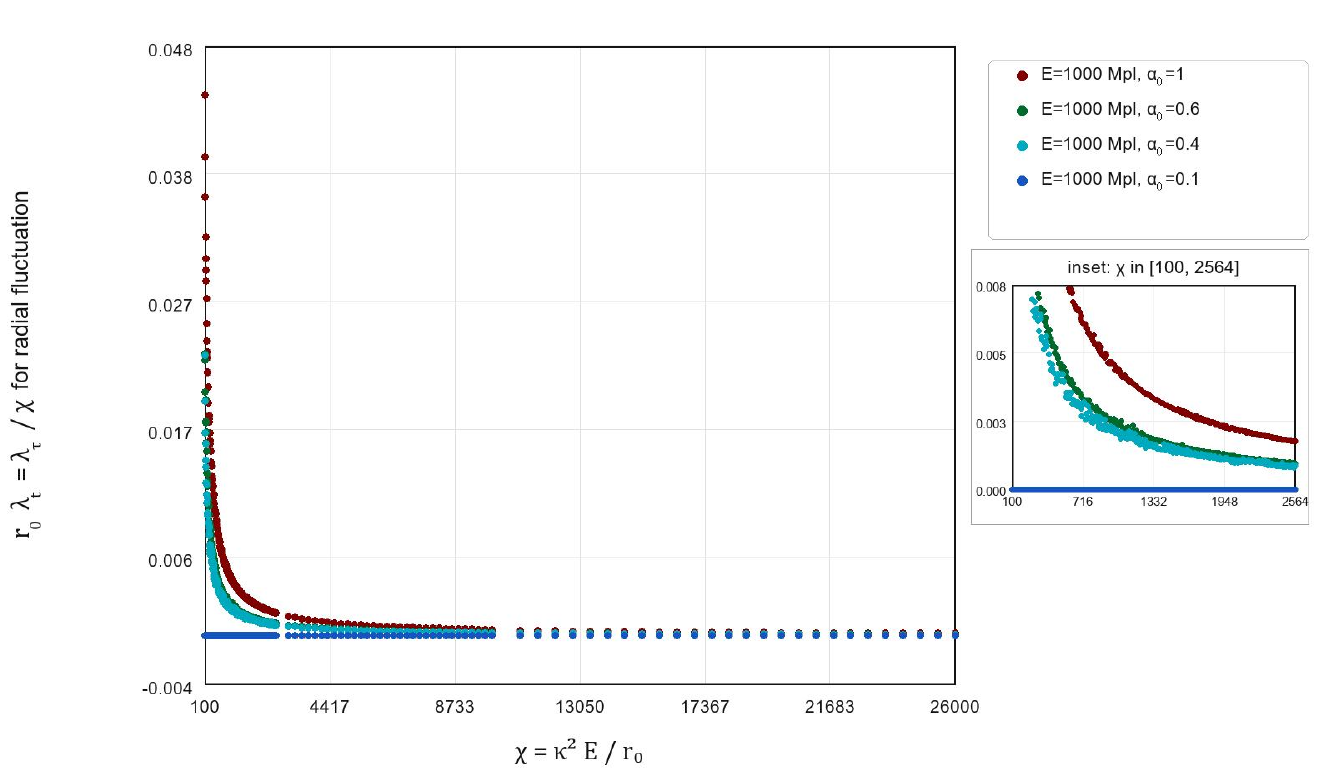}\\
    \includegraphics[scale=0.55]{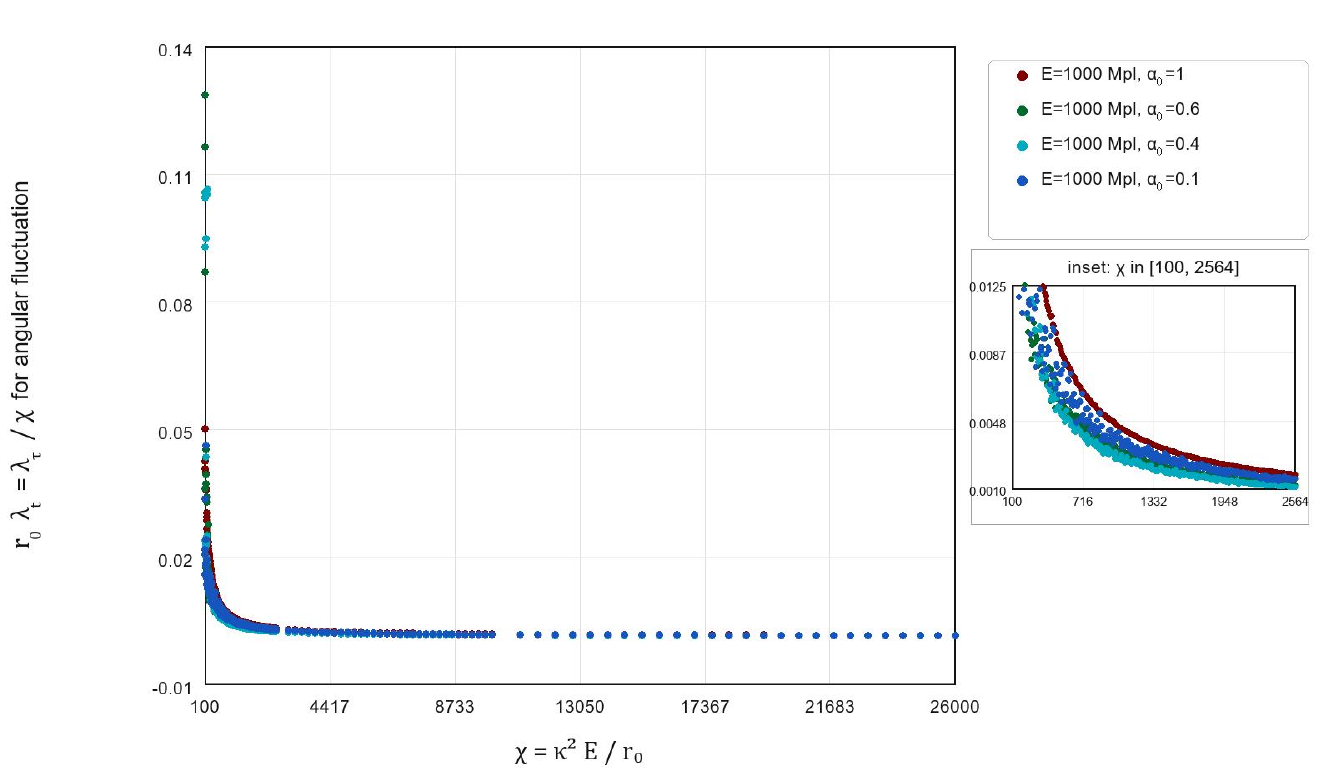}
 \caption{Dimensionless physical-time Lyapunov rate $r_0\lambda_t$ as a function of $\chi=\kappa^{2}E/r_0$ for the global-monopole wormhole. The upper and lower panels show the radial and angular fluctuation channels, respectively, while the different curves display the effect of the topological defect through representative values of $\alpha_0=\sqrt{1-\kappa^{2}\eta^{2}}$. The OTOC evolution, first-throat-crossing fit and conversion from $\lambda_\tau$ to $\lambda_t$ follow the same prescription as in the Ellis-Bronnikov case shown above. In the presence of the defect, the upper limit of the world-sheet mode-number sum is chosen to be the integer nearest to $\alpha_0\chi$, namely $n_{\mathrm{cutoff}}\simeq\alpha_0\chi$.}
 \label{LambdaVsChiAlpha0}
 \end{center}
 \end{figure}
\begin{figure}[H]
 \begin{center}
    \includegraphics[scale=0.55]{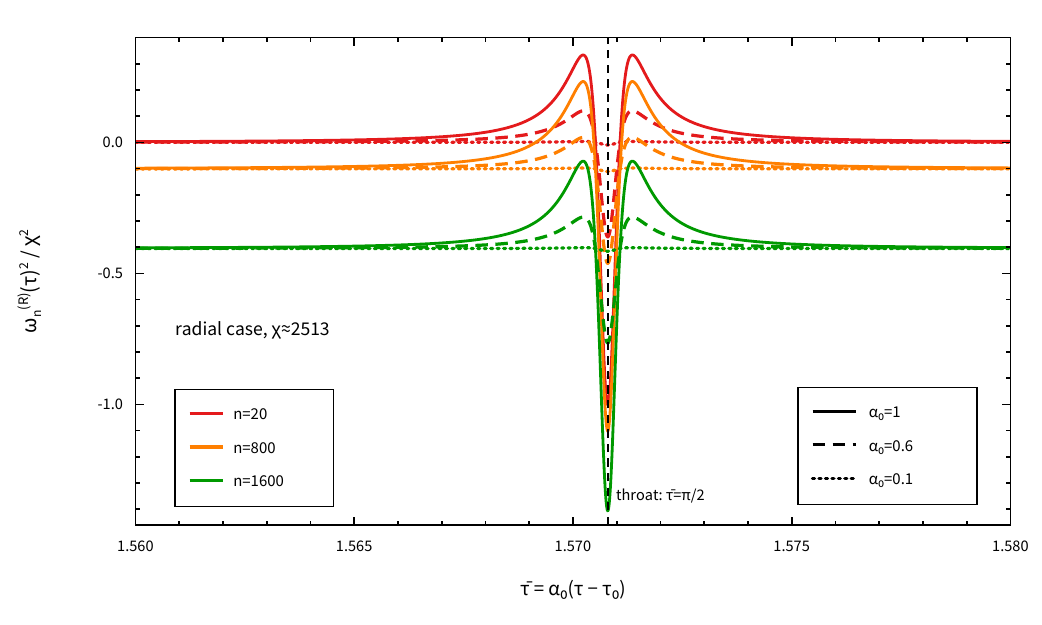}\\
    \includegraphics[scale=0.55]{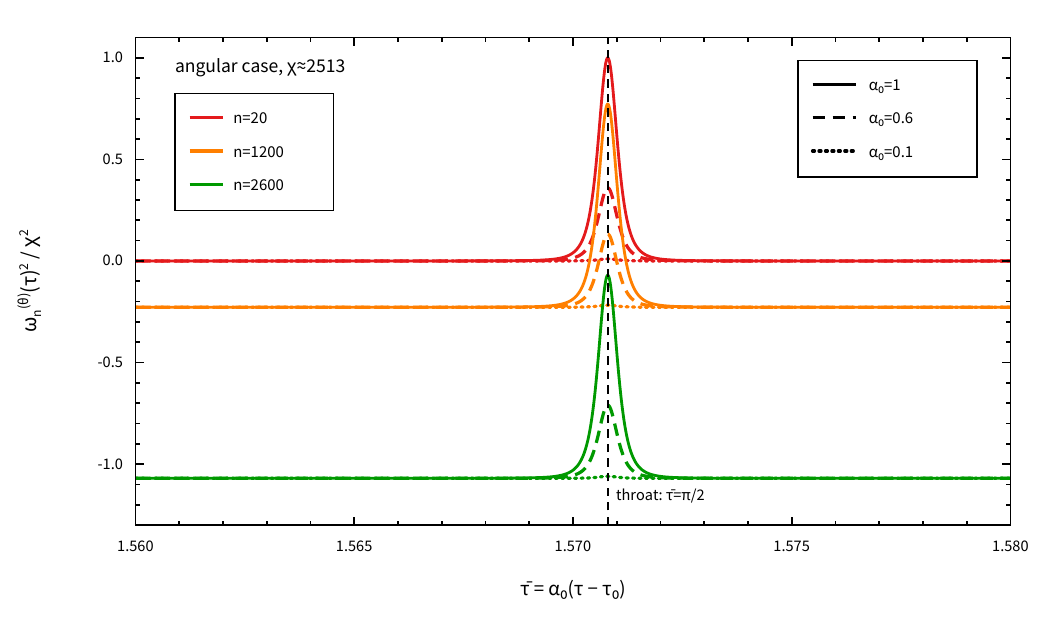}
 \caption{Dependence of the time-dependent effective frequency profiles on the deficit factor $\alpha_0$ for the global-monopole wormhole. The upper and lower panels show $\omega_{n}^{(\mathcal{R})}(\tau)^{2}/\chi^{2}$ and $\omega_{n}^{(\vartheta)}(\tau)^{2}/\chi^{2}$, respectively, for representative world-sheet mode numbers and several values of $\alpha_0$. The parameter $\chi\simeq2513$ is held fixed as in Fig.~\ref{EBWHEffOmegaSquare}, and the evolution is displayed in terms of the rescaled world-sheet time $\widetilde{\tau}=\alpha_0(\tau-\tau_0)$, for which the throat crossing occurs at $\widetilde{\tau}=\pi/2$. This figure extends the effective-frequency analysis of Fig.~\ref{EBWHEffOmegaSquare} and shows explicitly how the topological deficit modifies the height and profile of the throat-localized radial and angular potentials.}
 \label{FigGMWHEffOmega}
 \end{center}
 \end{figure}

The suppression caused by a larger deficit can also be seen directly at the level of the effective frequencies, without first extracting a Lyapunov exponent. Figure~\ref{FigGMWHEffOmega} compares the profiles in Eqs.~\eqref{EffOmegaSquareRadialGMWH} and \eqref{EffOmegaSquareAngularGMWH} at fixed $\chi$ for several values of $\alpha_0$. In the rescaled time variable used there, decreasing $\alpha_0$ lowers the throat-localized positive peaks and progressively closes the intervals with $\omega_n^2>0$, first for the higher modes and eventually for the low-lying modes as well. Since a mode can acquire exponential enhancement only while its effective frequency squared is positive, reducing both the height and the support of these intervals decreases the net amplification accumulated during a throat passage. The use of rescaled time in Fig.~\ref{FigGMWHEffOmega} isolates this deformation of the potential profile from the trivial stretching of the evolution in $\tau$; the Lyapunov analysis in Fig.~\ref{LambdaVsR0ForRadial} incorporates both effects. Taken together, the two figures identify the deficit angle as a control parameter for the probe-sector chaotic response: a sufficiently strong topological defect can substantially suppress, and in the radial channel nearly quench, the throat-induced OTOC growth, while the remaining non-monotonic features reflect the distinct curvature couplings of the two fluctuation polarizations.

 Combining the results obtained for the Ellis--Bronnikov and global-monopole wormholes, the numerical analysis in this section establishes two main conclusions. First, in the Ellis--Bronnikov geometry, the time-dependent fluctuation potentials generated during a throat passage produce transient mode amplification and an approximately linear growth regime in the logarithm of the OTOC amplitude. The corresponding nonzero effective Lyapunov rates in both fluctuation channels demonstrate that quantum-chaotic dynamics can be excited in the probe-string sector near the wormhole throat, even though the underlying circular-string trajectory is classically stable. Second, the global-monopole extension shows that this response is sensitive to the topology of the background: increasing the solid-angle deficit generally reduces the extracted value of $r_0\lambda_t$, narrows the set of locally amplifiable modes and, in the strong-defect regime, can nearly quench the radial OTOC growth. The wormhole throat therefore provides the local dynamical mechanism responsible for the amplification, while the deficit angle acts as a geometric control parameter governing its strength. Since these conclusions rely on numerical mode evolution, truncated mode sums and finite-window fits, the next subsection presents a set of complementary consistency checks based on Wronskian conservation, cutoff convergence, fit-window stability and the resulting error bars. These checks are logically separate from the main physical argument and are included to establish the numerical robustness of the results reported above.

\subsection{Numerical validation and uncertainty analysis\label{NumericalValidation}}

The physical discussion in Secs.~\ref{PureEBWHCase} and \ref{TopoGMWHCase} concerns the origin of the first-passage OTOC enhancement and its dependence on the throat scale and the solid-angle deficit. We now step aside from that main line of argument and perform an independent numerical audit of the calculation. The purpose is not to introduce another family of backgrounds or to extract an additional dynamical observable, but to test the successive numerical operations on which Figs.~\ref{CTvsTauAndLinearGrowth}--\ref{FigGMWHEffOmega} rely. Four logically distinct questions must be separated. First, does the time evolution of each mode preserve the canonical Wronskian normalization imposed in Eq.~\eqref{WronskianMode}, including for the high winding numbers that probe the narrow throat region? Second, are the coherently summed OTOC amplitude and the Lyapunov slope stable when the baseline mode cutoff is varied? Third, is the fitted slope supported by a finite neighborhood of admissible first-throat windows, rather than by one specially selected interval? Finally, how large is the combined uncertainty after the fit-window, cutoff and integration-resolution effects are propagated to the dimensionless rate $r_0\lambda_t$? To answer these questions without merely repeating the original numerical pipeline, we use an independent fourth-order Yoshida composition of exact free-oscillator drifts and time-dependent potential kicks, together with an adaptive phase grid that resolves the $O(\chi^{-1})$ structure around the throat. This hierarchy distinguishes an error in the mode evolution from sensitivity to the mode truncation and from the intrinsic ambiguity of extracting a finite-time growth rate.

For compactness, throughout this subsection $q_n(\tau)$ denotes either mode function: $q_n(\tau)=\mathcal{R}_n(\tau)$ in the radial channel and $q_n(\tau)=\vartheta_n(\tau)$ in the angular channel. Every occurrence of $q_n(\tau)$ below is to be understood separately for these two choices.

\subsubsection*{Wronskian preservation}

\begin{figure}[H]
 \begin{center}
    \includegraphics[scale=0.55]{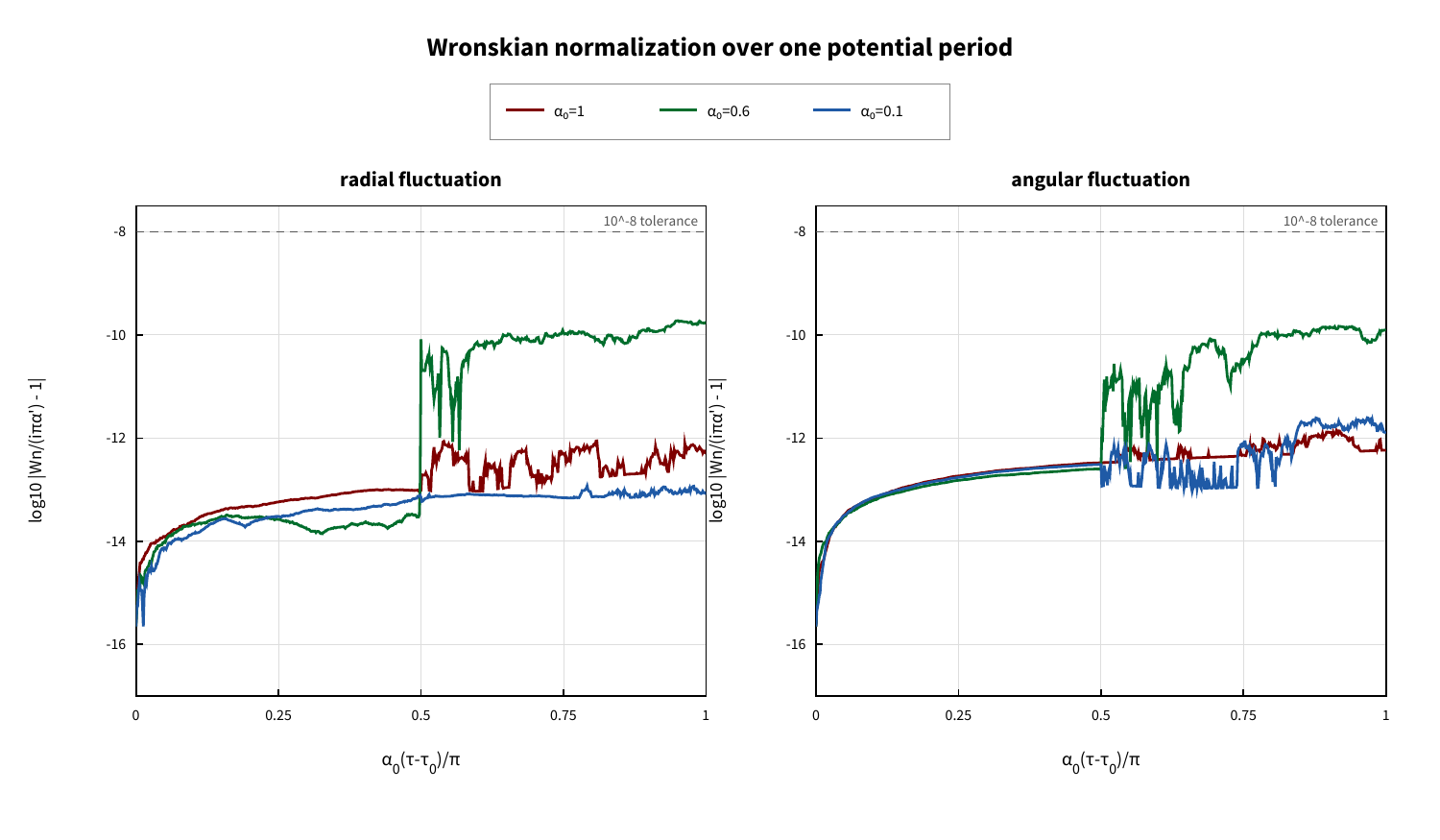}
 \caption{Conservation of the canonical Wronskian normalization during one period of the throat potential.  The left and right panels show the radial and angular fluctuation channels at $\chi=800\pi\simeq2513$, plotted against the normalized phase $\alpha_0(\tau-\tau_0)/\pi$.  The colors correspond to $\alpha_0=1,0.6$ and $0.1$.  At every time, each curve is the maximum of $\log_{10}|W_n/(i\pi\alpha')-1|$ over $n=(20,800,1600)$ for the radial modes and $n=(20,1200,2600)$ for the angular modes.  The horizontal dashed line denotes the conservative tolerance $10^{-8}$.  The maximum residual over all channels, modes and deficit factors is $1.88\times10^{-10}$, confirming that the initial normalization in Eq.~\eqref{InitialConditionModes} is preserved through the throat crossing.}
 \label{FigWronskiCheck}
 \end{center}
 \end{figure}

The mode equations have real time-dependent coefficients, and hence the exact evolution preserves
\begin{equation}
 W_n(\tau)=q_n(\tau)\dot q_n^*(\tau)-q_n^*(\tau)\dot q_n(\tau)=i\pi\alpha'.
 \label{WronskianConsistencyDefinition}
\end{equation}
Once the initial conditions in Eq.~\eqref{InitialConditionModes} are imposed. A drift of $W_n$ is therefore a direct, normalization-sensitive diagnostic of the numerical evolution; unlike a comparison of OTOC curves, it cannot be hidden by an overall amplitude rescaling. Figure~\ref{FigWronskiCheck} shows $\log_{10}|W_n/(i\pi\alpha')-1|$ at $\chi=800\pi\simeq2513$ over one period of the time-dependent potential. For each $\alpha_0$, the displayed curve is the largest residual among the representative modes $n=(20,800,1600)$ in the radial channel and $n=(20,1200,2600)$ in the angular channel. These choices simultaneously test a low mode, a mode inside or close to the amplifiable band, and a high rapidly oscillating mode. The largest residual found in the full scan is $1.88\times10^{-10}$, attained for $\alpha_0=0.6$, whereas most of the evolution remains at or below the $10^{-12}$ level. All six curves remain comfortably below the conservative $10^{-8}$ reference line. We therefore find no evidence for a loss of canonical normalization at the throat. Since the Yoshida evolution is symplectic by construction, however, Wronskian preservation alone does not bound accumulated phase or truncation errors; those independent effects are tested below through cutoff and resolution variations.

\subsubsection*{Mode-cutoff dependence}

\begin{figure}[H]
 \begin{center}
    \includegraphics[scale=0.55]{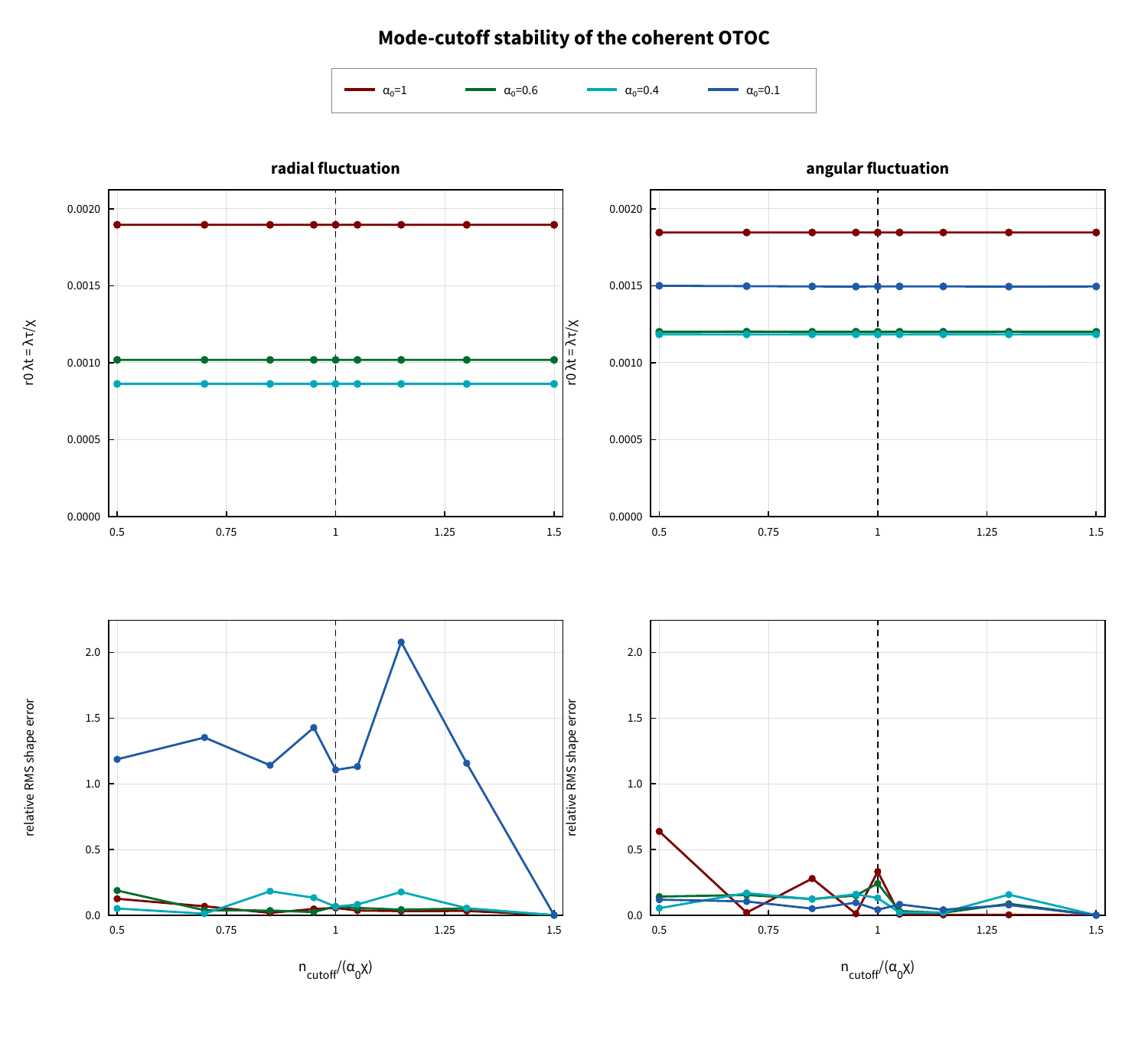}
 \caption{Mode-cutoff dependence of the coherently summed OTOC at $\chi=800\pi\simeq2513$. The left and right columns correspond to radial and angular fluctuations, and the colors denote $\alpha_0=1,0.6,0.4$ and $0.1$. The horizontal coordinate is the cutoff ratio $N/(\alpha_0\chi)$; the dashed vertical line marks the baseline prescription $N=N_{\rm phys}={\rm Round}(\alpha_0\chi)$. The upper panels show the dimensionless rate $r_0\lambda_t=\lambda_\tau/\chi$, evaluated for every cutoff with the same first-throat window selected at $N_{\rm phys}$, thereby isolating the cutoff effect. The absence of the radial $\alpha_0=0.1$ curve indicates that no accepted growth window exists, rather than a failed ODE solution. The lower panels show the relative RMS waveform error $\epsilon_{\rm shape}$ defined in Eq.~\eqref{CutoffShapeErrorDefinition}, using $N_{\rm ref}={\rm Round}(1.5\alpha_0\chi)$ as the reference; hence all lower curves vanish at $N/(\alpha_0\chi)=1.5$ by definition. The fitted rates are stable where growth is detected, while fine oscillatory structure can remain cutoff sensitive because of coherent cancellations.}
 \label{FigCutoffCheck}
 \end{center}
 \end{figure}

The OTOCs in Eqs.~\eqref{FinalOTOCamplitudeRadial} and \eqref{FinalOTOCamplitudeAngular} contain a coherent mode sum: the real mode contributions are summed before the modulus and logarithm are taken.  Writing the logarithmic amplitude schematically as
\begin{equation}
 g_N(\tau)=\log\left|\sum_{n=2}^{N}{\rm Re}\,q_n(\tau)\right|+\text{constant}
 \label{CutoffLogAmplitudeDefinition}
\end{equation}
makes clear that changing $N$ can alter both the smooth growth envelope and the positions of destructive-interference nodes. We test these two effects separately at $\chi=800\pi$ for $\alpha_0=1,0.6,0.4$ and $0.1$. The baseline truncation used in Secs.~\ref{PureEBWHCase} and \ref{TopoGMWHCase} is denoted by $N_{\rm phys}={\rm Round}(\alpha_0\chi)$, with the Ellis--Bronnikov value recovered at $\alpha_0=1$. This notation labels the mode-counting prescription motivated by the amplifiable band; it should not be interpreted as a fundamental ultraviolet cutoff of the string theory. In Fig.~\ref{FigCutoffCheck} the cutoff is varied from $0.5N_{\rm phys}$ to $1.5N_{\rm phys}$. The upper panels show $r_0\lambda_t$ obtained by keeping fixed the best window selected at $N=N_{\rm phys}$. This is essential: allowing the window to change simultaneously would mix cutoff dependence with fit-selection dependence. Whenever an admissible growth window exists, the upper curves are nearly horizontal. In the narrower $\pm10\%$ variation used below for the error budget, the largest relative cutoff contribution to the rate is only $1.13\times10^{-3}$.

The lower panels test the full waveform rather than only its fitted slope.  With $\Delta g_N(\tau)=g_N(\tau)-g_N(\tau_{\rm initial})$ and $N_{\rm ref}={\rm Round}(1.5\alpha_0\chi)$, the plotted quantity is
\begin{equation}
 \epsilon_{\rm shape}(N)=
 \frac{\left\langle\left[\Delta g_N(\tau)-\Delta g_{N_{\rm ref}}(\tau)\right]^2\right\rangle^{1/2}}
 {\left\langle\Delta g_{N_{\rm ref}}(\tau)^2\right\rangle^{1/2}}\,.
 \label{CutoffShapeErrorDefinition}
\end{equation}
Consequently, every curve vanishes at $N/N_{\rm phys}=1.5$ by construction, and convergence need not be monotonic because the sum in Eq.~\eqref{CutoffLogAmplitudeDefinition} is coherent rather than positive term by term. The distinction between slope stability and pointwise waveform stability is physically relevant. The former supports the Lyapunov trends reported in Secs.~\ref{PureEBWHCase} and \ref{TopoGMWHCase}, whereas the latter warns that the phase of the fine OTOC oscillations should not be assigned cutoff-independent significance.

The radial curve at $\alpha_0=0.1$ provides the sharpest example of this distinction. No cutoff in the scan yields a first-throat interval that passes the adopted exponential-growth criteria, so the upper-left panel contains no corresponding blue rate. The coherent sum is instead dominated by oscillatory contributions; destructive interference repeatedly drives it close to zero, a small change of $N$ shifts these nodes, and the logarithm magnifies that displacement into a large, non-monotonic shape error. This is not a failure of the mode solver. For this case the normalized Wronskian residual is $1.18\times10^{-13}$, and two successive step refinements leave the large shape error unchanged within the expected convergence accuracy. The correct conclusion is therefore that the detailed radial waveform in this strong-defect, non-growing regime is mode-cutoff sensitive. It does not supply a cutoff-stable Lyapunov exponent, consistently with assigning no detected radial growth rate there; by contrast, the fitted rates in the regimes with an accepted growing window are robust under the stated cutoff variations.

\subsubsection*{Fit-window stability}

\begin{figure}[H]
 \begin{center}
\includegraphics[scale=0.55]{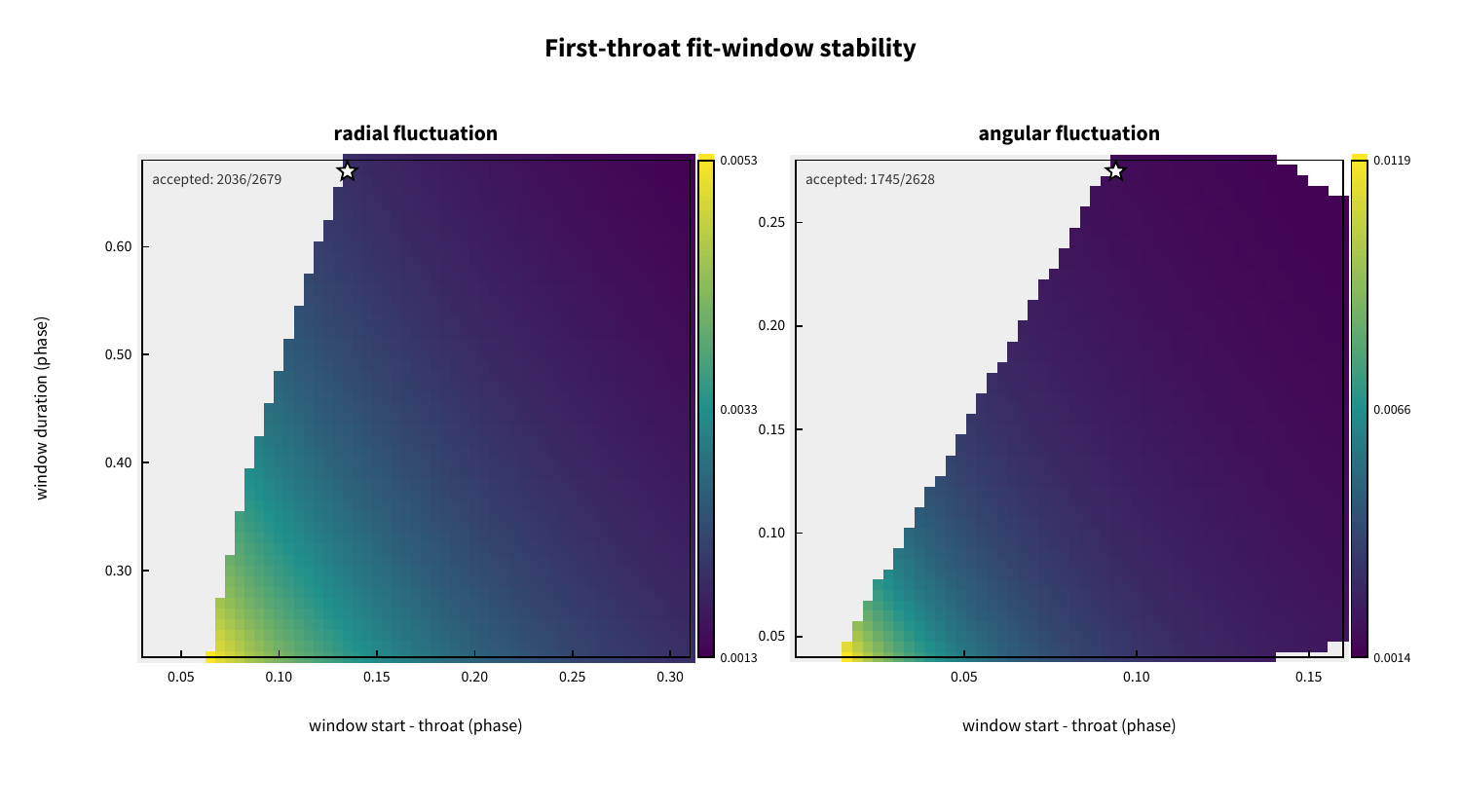}
 \caption{Stability of the first-throat Lyapunov fit under changes of its starting point and duration, evaluated at $\chi=800\pi$, $\alpha_0=1$ and $N=N_{\rm phys}=2513$. The left and right heat maps show the radial and angular channels. Their axes give the fit-window start relative to the throat and the window duration in the phase variable $\alpha_0(\tau-\tau_0)$. Colored cells pass all channel-dependent positivity, net-growth and $R^2$ cuts specified in the text, and the color gives the resulting $r_0\lambda_t$; gray cells are rejected. The radial scan accepts $2036/2679$ candidate windows and the angular scan accepts $1745/2628$. The white star identifies the maximum-score window used for the central rate. The extended accepted regions show that the selected first-throat growth window is not isolated within the stated criteria, while their finite color spread quantifies the fit-window systematic.}
 \label{FigFitWindowCheck}
 \end{center}
 \end{figure}

The exponent extracted from a finite throat passage is not an asymptotic Lyapunov exponent, and its dominant systematic uncertainty is the choice of the interval over which $g_N(\tau)$ is approximated by a straight line.  Figure~\ref{FigFitWindowCheck} therefore scans the starting offset from the first throat and the duration of the fit at the reference point $\chi=800\pi$, $\alpha_0=1$ and $N=2513$.  For the radial channel the phase offset is varied over $0.03$--$0.31$ in steps of $0.005$, and the duration over $0.22$--$0.68$ in steps of $0.01$.  A radial window is accepted only if $\lambda_\tau\geq0.05$, the net logarithmic increase is at least $0.8$, and $R^2\geq0.95$.  Because the angular enhancement is shorter, its offset is scanned over $0.001$--$0.16$ in steps of $0.003$ and its duration over $0.04$--$0.28$ in steps of $0.005$, with acceptance conditions $\lambda_\tau\geq0.05$, logarithmic increase at least $0.25$, and $R^2\geq0.97$.  Windows extending beyond the prescribed first-passage domain are discarded.

The heat maps display $r_0\lambda_t$ for every accepted window; gray cells fail at least one of the slope, growth or linearity requirements. The accepted sets form connected two-dimensional regions, containing $2036$ of $2679$ radial candidates and $1745$ of $2628$ angular candidates. Thus, within the stated acceptance criteria, the linear-growth diagnosis is not supported by a single isolated interval. The white star marks the window that maximizes $\Delta g\,R^2\sqrt{\Delta\tau}$ and supplies the central value used elsewhere, namely $r_0\lambda_t=1.8960\times10^{-3}$ for the radial channel and $1.8460\times10^{-3}$ for the angular channel. In both panels the selected window lies near the upper boundary of the allowed duration. This is consistent with the preference of the score for larger net growth and longer intervals, together with the restriction to the first-passage domain, and the selected point should therefore not be interpreted as an unconstrained interior optimum. At the same time, the color variation across the accepted domains shows that the slope is not exactly window independent, especially for the shorter angular episode. We treat this spread as a finite-time systematic uncertainty rather than concealing it through a single best-fit error returned by linear regression.

\subsubsection*{Uncertainty budget for the Lyapunov rates}
\begin{figure}[H]
 \begin{center}
    \includegraphics[scale=0.55]{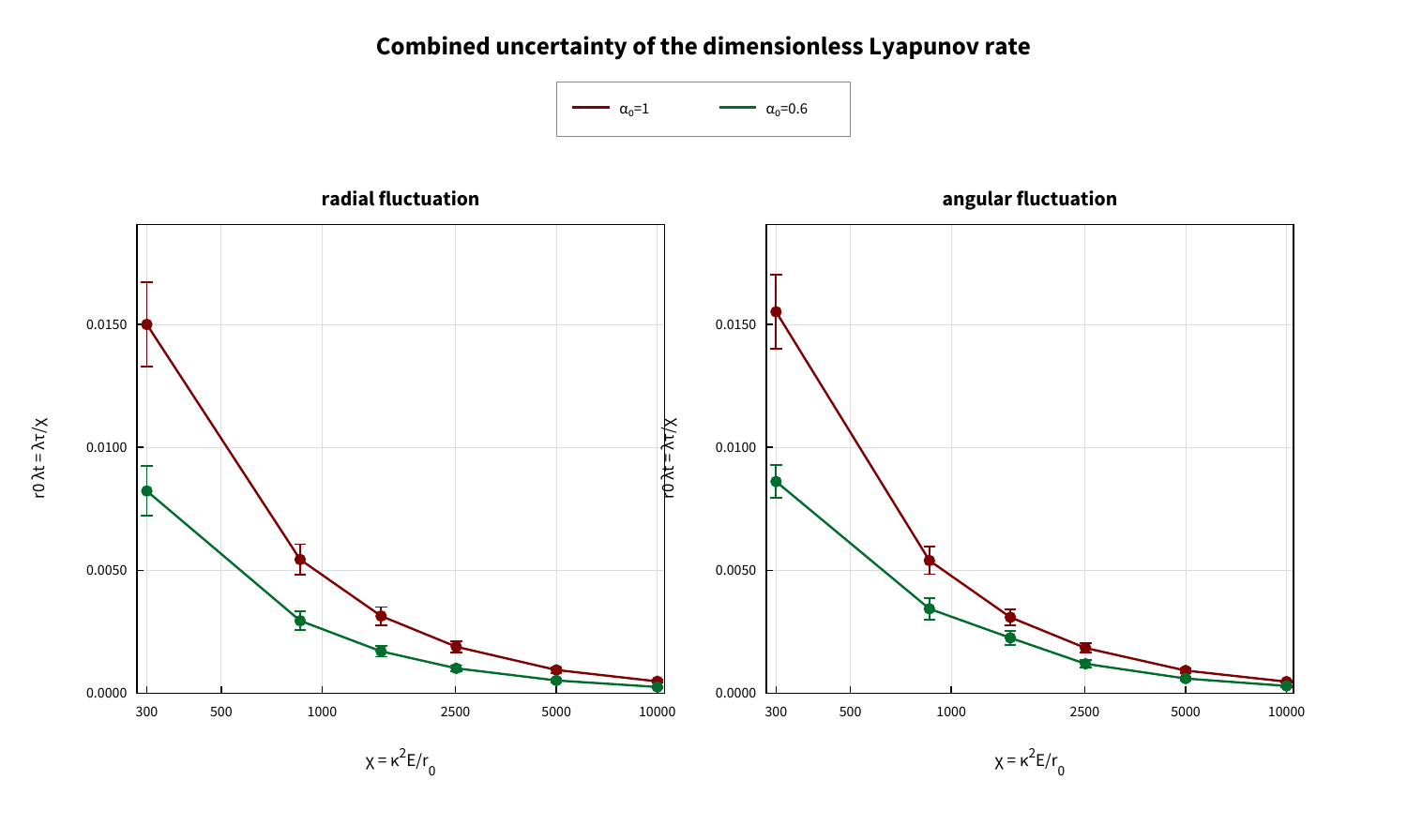}
 \caption{Combined numerical uncertainty of the dimensionless physical-time Lyapunov rate $r_0\lambda_t=\lambda_\tau/\chi$. The left and right panels show the radial and angular fluctuation channels for $\alpha_0=1$ and $0.6$ at $\chi=300,860,1500,800\pi,5000$ and $10000$; connecting lines are included only to guide the eye. Central values use the maximum-score first-throat fit at the baseline cutoff $N_{\rm phys}={\rm Round}(\alpha_0\chi)$. Each vertical error bar is the quadrature sum in Eq.~\eqref{TotalLyapunovNumericalError} of the nearby-window 16th--84th percentile half-width, the half-range from cutoff factors $0.9,1.0$ and $1.1$ evaluated in a fixed window, and the change obtained after doubling the integration resolution. Over the sampled points, the total relative uncertainty is about $10.3\%$--$13.5\%$ and is dominated by the finite-window contribution; the cutoff and integration contributions reach at most $0.113\%$ and $0.0071\%$, respectively.}
 \label{FigErrorBarsCheck}
 \end{center}
 \end{figure}

We finally propagate the preceding tests into an uncertainty on the dimensionless physical-time rate. Figure~\ref{FigErrorBarsCheck} uses representative values $\alpha_0=1$ and $0.6$ in both channels and $\chi=300,860,1500,800\pi,5000$ and $10000$. The central value at each point is obtained from the maximum-score first-throat window at the baseline cutoff $N_{\rm phys}={\rm Round}(\alpha_0\chi)$. Three numerical contributions are then estimated separately. The window uncertainty is one half of the 16th--84th percentile interval of the accepted rates in a neighborhood of the selected window, allowing its starting offset to vary by at most $\max(0.02,0.2\Delta\widetilde{\tau})$ and its duration by $0.2\Delta\widetilde{\tau}$. The cutoff uncertainty is one half of the range obtained from $0.9N_{\rm phys}$, $N_{\rm phys}$ and $1.1N_{\rm phys}$ while holding the baseline window fixed. The integration uncertainty is the absolute shift produced by halving both the coarse phase step and the local throat-resolution scale. Treating these contributions as independent for the purpose of the error budget, we combine them in quadrature,
\begin{equation}
 \sigma_{\rm tot}=
 \sqrt{\sigma_{\rm window}^{2}+\sigma_{\rm cutoff}^{2}+\sigma_{\rm int}^{2}}\,.
 \label{TotalLyapunovNumericalError}
\end{equation}

The resulting relative uncertainties lie between approximately $10.3\%$ and $13.5\%$ over the displayed samples. They are overwhelmingly dominated by the fit-window ensemble: the largest relative cutoff term is $0.113\%$, while the largest integration-resolution term is $0.0071\%$ in the radial channel and $1.6\times10^{-7}\%$ in the angular channel. Thus, at the sampled parameter values, neither ODE discretization nor a modest displacement of the baseline cutoff can account for the Lyapunov trends reported in Secs.~\ref{PureEBWHCase} and \ref{TopoGMWHCase}. The error bars instead measure the unavoidable ambiguity of assigning one slope to a transient growth episode. The overall decrease of $r_0\lambda_t$ with $\chi$ and its suppression by the deficit are resolved at these representative points, whereas the checks do not certify every plotted point and structures smaller than the quoted window uncertainty should not be overinterpreted. They support the numerical credibility of the finite-time Lyapunov analysis without promoting the fitted rates to asymptotic chaos exponents.

\section{Conclusion and discussion \label{ConcluDiscuss}}

\subsection{Conclusion}

In this work we have used a circular string to probe the quantum response of traversable wormhole throats in the Ellis--Bronnikov geometry and its global-monopole extension. The circular ansatz admits an exactly solvable periodic radial orbit with no exponentially growing radial variational mode. Around this embedding, we derived the covariant quadratic action for the two physical transverse polarizations, imposed the canonical Wronskian normalization and expressed their unequal-time commutator amplitudes in terms of the corresponding mode functions. This construction connects the throat geometry to a quantum dynamical observable without introducing an event horizon or an externally supported unstable particle orbit. Its central outcome is a throat-associated enhancement of the OTOC amplitudes whose strength depends on both the probe energy relative to the throat scale and the topological defect.

For the Ellis--Bronnikov wormhole, Figs.~\ref{EBWHEffOmegaSquare}--\ref{LambdaVsR0ForRadial} display the successive stages of the analysis. The effective-frequency and mode-function profiles in Fig.~\ref{EBWHEffOmegaSquare} distinguish the two polarizations: the angular channel can be locally amplified at the throat, whereas the radial effective frequency squared is negative at the exact crossing and its enhancement is accumulated through the surrounding region. Fig.~\ref{CTvsTauAndLinearGrowth} then shows finite intervals of approximately exponential OTOC growth associated with the first throat passage, and Fig.~\ref{LambdaVsR0ForRadial} summarizes the corresponding dimensionless physical-time rates $r_0\lambda_t=\lambda_\tau/\chi$, with $\chi=\kappa^{2}E/r_0$. This quantity measures the Lyapunov growth accumulated over the characteristic throat timescale; within the parameter range and first-passage prescription considered, it is positive in both channels and decreases as $\chi$ increases. A stronger instantaneous tidal contribution therefore does not necessarily imply faster accumulated OTOC growth, which depends on the complete evolution through the time-dependent potential. Figs.~\ref{LambdaVsChiAlpha0} and \ref{FigGMWHEffOmega} extend this analysis to the global-monopole wormhole and demonstrate that the response can be tuned by the solid-angle deficit. As shown by the extracted rates in Fig.~\ref{LambdaVsChiAlpha0}, increasing the deficit, and hence decreasing $\alpha_0$, suppresses the dimensionless radial growth rate at fixed $\chi$, complementing its decrease with increasing $\chi$. Sufficiently strong defects can also remove an identifiable first-throat growth window under the adopted extraction criteria. The angular channel is likewise suppressed at moderate and large $\chi$, although it retains a non-monotonic dependence on the defect at small $\chi$. The effective-frequency profiles in Fig.~\ref{FigGMWHEffOmega} relate these trends to the deformation of the radial potential and to the reduction of the angular local-amplification band, $n^2<1+(\chi^2-1)\alpha_0^2$. The defect therefore modifies the dynamical response in a polarization-dependent manner rather than merely rescaling a universal Lyapunov exponent. This behavior is qualitatively consistent with the suppression of probe-string entanglement by a larger deficit found in Ref.~\cite{Li:2026rut}, while providing a distinct characterization through unequal-time commutators.

The independent numerical audit in Sec.~\ref{NumericalValidation} supports the computational pipeline and interpretation underlying Figs.~\ref{EBWHEffOmegaSquare}--\ref{FigGMWHEffOmega}, particularly the quantitative conclusions drawn from Figs.~\ref{CTvsTauAndLinearGrowth}--\ref{LambdaVsChiAlpha0}. Wronskian preservation, variations of the mode cutoff and integration resolution, and fit-window scans support the stability of the principal trends within the estimated numerical uncertainties. The decrease with $\chi$ and the overall defect-induced suppression are resolved at the tested parameter points, whereas fine point-to-point structure remains inconclusive when comparable to the fit-window uncertainty.

The physical significance of these results is that a regular, horizonless throat can produce a finite-time OTOC signal of the type commonly associated with quantum chaos, even when the circular radial motion is integrable. In this respect, the present construction connects two complementary lines of investigation. Reference~\cite{Hashimoto:2016dfz} identified a universal local instability for an externally supported particle near a non-extremal black-hole horizon, with a Lyapunov exponent set by the surface gravity, whereas Ref.~\cite{Hashimoto:2017oit} formulated OTOCs directly in quantum mechanics and also illustrated that their relation to classical chaos need not be one-to-one. Our analysis implements the corresponding semiclassical question for an extended probe in a traversable-wormhole background: the horizon and externally generated unstable equilibrium are replaced by the time-dependent tidal potentials experienced by the string fluctuations during a passage through the regular throat, while unequal-time commutators quantify the resulting quantum dynamical sensitivity. It thereby supplies a horizonless gravitational setting in which OTOC growth can be studied mode by mode and its dependence on both the throat scale and a topological defect can be resolved. The comparison must, however, distinguish the degrees of freedom involved: radial stability within the circular ansatz does not establish stability against all transverse deformations. Moreover, at quadratic order the Heisenberg equations are linear and the commutators are governed by the same evolution matrices as the corresponding classical fluctuation equations. Exponential OTOC growth can therefore encode local instability or squeezing without establishing many-body chaos or thermalization, as emphasized more generally in Refs.~\cite{Hashimoto:2017oit,Xu:2019lhc}. Our effective Lyapunov exponents characterize this finite-time dynamical sensitivity of the quantized transverse sector; they should not be identified with asymptotic chaos exponents of the full interacting string theory or interpreted as evidence that quantization alone creates an instability absent from the same classical fluctuation problem.

\subsection{Discussion}

The preceding conclusions are deliberately limited to finite-time OTOC growth in the quadratic probe sector. Several extensions can test which features persist beyond this controlled approximation:

\begin{itemize}
\item \textbf{Numerical systematics and repeated passages.} The audit in Sec.~\ref{NumericalValidation} could be strengthened by comparing independent solvers and cutoff prescriptions and by propagating fit-window uncertainties coherently across the full parameter scan. Following several throat crossings and applying Floquet analysis to the periodic embedding \cite{Nizami:2020agu} would further distinguish a transient first-passage response from a persistent instability.

\item \textbf{Nonlinear world-sheet dynamics.} The two transverse polarizations exhaust the local physical string fluctuations in four spacetime dimensions \cite{Larsen:1993mx,Garriga:1991ts,Guven:1993ex}, but a complete stability analysis must relax the circular background ansatz while enforcing the world-sheet constraints. Cubic and quartic interactions, guided by nonlinear effective-string expansions and higher-order world-sheet Hamiltonians \cite{Aharony:2013LongStrings,Astolfi:2010QuantumStrings}, would couple different modes and test whether the observed amplification develops into sustained information transfer, non-Gaussian correlations and eventual OTOC saturation.

\item \textbf{Embedding in interacting string theory.} A full treatment requires a wormhole embedded in a consistent string background, control over both $\alpha'$ and string-loop corrections, and the backreaction on the geometry and its supporting matter. Covariant semiclassical superstring quantization and string corrections to scrambling provide useful benchmarks \cite{Drukker:2000SemiclassicalString,McLoughlin:2010QuantumStrings,Shenker:2014Stringy,Banerjee:2018Strings}. Holographic OTOCs in AdS traversable wormholes and coupled-SYK wormhole phases \cite{Cubrovic:2021puw,Nosaka:2020nuk} offer complementary interacting comparisons, although they probe boundary many-body observables rather than the local commutator studied here.

\item \textbf{Open-system effects.} Selected fluctuation modes may instead be retained as a subsystem whose reduced density matrix is obtained by tracing out other modes or environmental fields. Influence-functional and open effective field theory methods \cite{Hu:2008StochasticGravity,Burgess:2014OpenEFT} could determine how noise, dissipation and decoherence modify the unequal-time commutators. A Lindblad description is appropriate only under controlled weak-coupling and Markovian assumptions; otherwise memory effects must be retained. The multitime correlators must also be evolved explicitly, since the reduced density matrix alone does not determine an OTOC.

\item \textbf{Alternative probes and diagnostics.} Quantizing the externally supported particle of Ref.~\cite{Hashimoto:2016dfz} with curved-space worldline methods \cite{Bastianelli:2002Worldline,Bastianelli:2005Worldline} would permit a direct comparison while retaining the original force, quantum state and boundary conditions. Canonical quantization of the double-rod pendulum provides a finite-dimensional benchmark relating OTOCs to level statistics and complexity \cite{Sun:2023DoublePendulum}, while quantum fluctuations of a moving D3-brane connect squeezed-state circuit complexity with chaotic evolution for an extended probe \cite{Li:2021MovingD3}.

\item \textbf{Other horizonless geometries.} Useful backgrounds include static wormholes that closely mimic black holes \cite{Damour:2007Foils}, rotating traversable wormholes \cite{Teo:1998Rotating}, vector--tensor Morris--Thorne wormholes \cite{Li:2020VectorTensorWormhole}, evolving thin-shell throats \cite{Li:2018EvolvingThinShell}, oscillating boson stars \cite{Yang:2025BosonLensing} and layered compact objects \cite{Su:2026LayeredEchoes}. AdS-soliton instabilities \cite{Cai:2011MagneticSoliton,Cai:2012SolitonEntanglement}, analogue-gravity mappings \cite{Yang:2019CurvedSpacetime}, classical lensing diagnostics \cite{Tsukamoto:2017Microlensing} and domain-wall-generated wormholes \cite{Deng:2016DomainWalls} provide complementary tests of how geometry, observables and throat dynamics affect the OTOC response.

\item \textbf{Quantum-supported throats and topology.} Further benchmarks include one-loop backreaction on topological wormholes \cite{Mehulic:2026Backreaction}, semiclassical monopole wormholes \cite{Rahaman:2023SemiclassicalMonopoles}, fermionic vacuum polarization near a monopole throat \cite{Li:2026FermionPolarization} and magnetically charged wormholes supported by quantum matter \cite{Maldacena:2018FourDimensions}. These constructions involve different notions of topology and charge, and vacuum polarization on a fixed background does not by itself define a self-consistent corrected geometry. Topological censorship \cite{Friedman:1993TopologicalCensorship} constrains causal access to topology under specific global assumptions but supplies no Lyapunov bound. Any relation to the defect-induced suppression found here would therefore require tracking the backreacted causal structure together with the OTOCs.
\end{itemize}

\section{Acknowledge}
Ai-chen Li was supported by funding from the China Scholarship Council (CSC) with  Grant No.202008620074. Xin-Fei Li is supported by NSFC with Grant No. 12565008, Youth Program of Natural Science Foundation of Guangxi with Grant No. 2021GXNSFBA075049 and Doctor Start-up Foundation of Guangxi University of Science and Technology with Grant No.19Z21.

\bibliographystyle{JHEP}
\bibliography{bibliography}

\end{document}